\documentclass[11pt,a4paper,reqno]{amsart}
\usepackage[foot]{amsaddr}
\usepackage[utf8]{inputenc}
\usepackage{color}
\usepackage{amsmath}
\usepackage{amssymb}
\usepackage{bm}
\usepackage{bbm}
\usepackage{hyperref}
\usepackage{amsthm}
\usepackage[dvipsnames]{xcolor}
\usepackage{graphicx}
\usepackage{diagbox}

\usepackage[shortlabels]{enumitem}

\usepackage{caption}
\usepackage{subcaption}
\usepackage{siunitx}

\usepackage{algorithm}
\usepackage{algpseudocode}
\usepackage[normalem]{ulem}

\theoremstyle{plain}

\theoremstyle{definition}

\theoremstyle{remark}

\numberwithin{equation}{section}

\makeatletter
\def\th@plain{%
  \thm@notefont{}% same as heading font
  \itshape % body font
}
\def\th@definition{%
  \thm@notefont{}% same as heading font
  \normalfont % body font
}
\makeatother

\usepackage{geometry}

\usepackage[english]{babel}

\usepackage[
backend=biber,
style=numeric-comp,
sorting=none,
giveninits=true
]{biblatex}

\bibliography{papers.bib}

\graphicspath{{./figures/}}

\newcommand{\La}{\Lambda}
\newcommand{\R}{\mathbb{R}}

\newcommand{\N}{\mathbb{N}}

\DeclareMathOperator*{\argmin}{arg\,min}

\title[Anisotropic Power Diagrams and polycrystals]{Anisotropic power diagrams for polycrystal modelling: Efficient generation of curved grains via optimal transport}
\author{M.~Buze$^1$}
\address{$^1$Maxwell Institute for Mathematical Sciences and Department of Mathematics, Heriot-Watt University, Edinburgh, EH14 4AS, United Kingdom
}
\email{m.buze@hw.ac.uk,d.bourne@hw.ac.uk}
\author{J.~Feydy$^2$}
\address{$^2$ HeKA team, INRIA Paris, INSERM, Université Paris-Cité, 2-10 rue d'Oradour-sur-Glane, Paris, France}
\email{jean.feydy@inria.fr}
\author{S.M.~Roper$^3$}
\address{$^3$ School of Mathematics and Statistics, University of Glasgow, University Avenue, Glasgow G12 8QQ, United Kingdom}
\email{steven.roper@glasgow.ac.uk}
\author{K.~Sedighiani$^4$}
\address{$^4$ Tata Steel, R\&D, IJmuiden, Netherlands}
\email{karo.sedighiani@tatasteeleurope.com}
\author{D.P.~Bourne$^1$}
\date{\today}
\begin{document}

\maketitle

\begin{abstract}
    The microstructure of metals and foams can be effectively modelled with anisotropic power diagrams (APDs), which provide control over the shape 
    of individual grains. One major obstacle to the wider adoption of APDs is the computational cost that is associated with their generation. We propose a novel approach to generate APDs with prescribed statistical properties, including fine control over the size of individual grains. To this end, we rely on fast optimal transport algorithms that stream well on Graphics Processing Units (GPU) and handle non-uniform, anisotropic distance functions. This allows us to find large APDs that best fit experimental data and generate synthetic high-resolution microstructures in (tens of) seconds. This unlocks their use for computational homogenisation, which is especially relevant to machine learning methods that require the generation of large collections of representative microstructures as training data. The paper is accompanied by a Python library, \textsc{PyAPD}, which is freely available at: \url{www.github.com/mbuze/PyAPD}.
\end{abstract}

%\noindent \cdb{Example comment.} {\db example change}\\
%\csr{Example comment.} {\sr example change}\\
%\cjf{Example comment.} {\jf example change}\\
%\cks{Example comment.} {\ks example change}\\

\begin{figure}[h!]
    \centering
    \begin{subfigure}[c]{0.3\textwidth}
        \centering
        \subcaption{}
        \includegraphics[width=\textwidth]{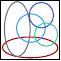}
        %\caption{EBSD scan}
        %\label{fig:y equals x}
    \end{subfigure}
    \hfill
    \begin{subfigure}[c]{0.3\textwidth}
        \centering
        \subcaption{}
        \includegraphics[width=\textwidth]{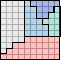}
        %\caption{EBSD scan}
        %\label{fig:y equals x}
    \end{subfigure}
    \hfill
    \begin{subfigure}[c]{0.3\textwidth}
        \centering
        \subcaption{}
        \includegraphics[width=\textwidth]{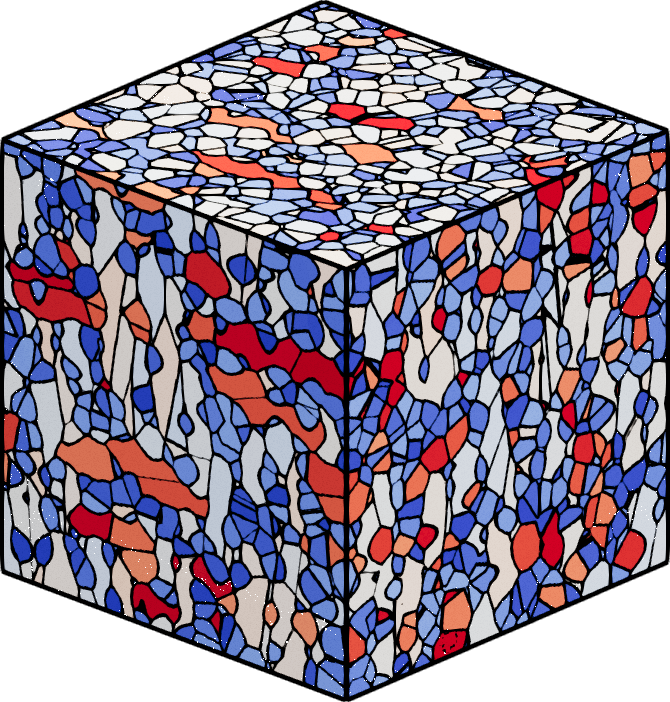}
        %\caption{EBSD scan}
        %\label{fig:y equals x}
    \end{subfigure}
    \caption{(A): We model grains in polycrystalline metals as anisotropic power cells, with desired centroid positions, volumes and shapes specified by ellipsoids.  
    %{\db \sout{ellipsoids}} {\db anisotropic power cells} \cdb{I wouldn't say that we model the grains as ellipsoids. We model them as APD cells, which need not look like ellipsoids?} with desired centroid positions, {\db \sout{$x_i$}} \cdb{This isn't quite accurate. The $x_i$ are the seeds, not the centroids. In fact, for the case $\mathbf{A}_i=I$ for all $i$, you can uniformly translate and dilate all the seeds, $x_i \mapsto \lambda x_i + t$, without changing the cells (and hence the centroids), provided that you change the weights appropriately.} 
    %\cmb{hence it being preceded by "desired"? In a way it is where we want the centroid to be and this is approximately the case when ellipses nicely fill the space, and also the point of us using Lloyds algorithm?} 
    %volumes, {\db \sout{$v_i$}} and {\db shapes specified by ellipsoids} {\db \sout{anisotropy matrices $\mathbf{A}_i$}}. \cdb{I'm suggesting removing the notation $x_i,v_i,\mathbf{A}_i$  to avoid introducing notation in the visual abstract.} 
    Accommodating these constraints in a 2D or 3D spatial domain is a difficult space-filling problem. %{\db \sout{packing}} {\db space-filling} problem. %\cdb{I don't think we solve any packing problem. I think we should be more precise in the visual abstract.} 
    %\cmb{This was written by Jean so perhaps he can comment. I get your point, but I think it is tricky to have some balance here, because I see the role of the visual abstract as giving readers some intuition with as few technicalities as possible and I am happy to sacrifice preciseness for it -- visually it is great to have ellipses of different sizes and shapes which are overlapping, it is an intuitive concept and nicely captures all the data we supply to each cell of the APD without ever requiring the reader to know what an "anisotropic power cell" is (which I guess you mean as "a cell in an anisotropic power diagram? Because not sure if one can define an anisotropic power cell separate without a reference to the APD it belongs to? Also, I agree that strictly speaking we do not solve a packing problem, but it is the overlap between ellipses that we see in panel (A) that leads to this problem being hard, we need to somehow "stretch" these ellipses and "pack" them into the domain. Perhaps we solve a "space filling" problem? } 
    (B): We use a semi-discrete optimal transport solver to find a tessellation of the spatial domain that satisfies these constraints approximately. Using a pixel grid that is fine enough, we can enforce an arbitrary tolerance on the volumes. %{\db \sout{$v_i$.}} 
    Unlike previous approaches, our method can handle both isotropic cells (blue and green) and elongated crystals (red, grey). 
    (C): Our method streams well on Graphics Processing Units (GPUs), allowing us to generate synthetic high-resolution 3D microstructures in seconds.}
       \label{fig:visual-abstract}
\end{figure}

\section{Introduction}

%\cdb{We should spellcheck the document; the word anisotropic is misspelled in several places, including the short title.} \cmb{Indeed, and other thing like that is spelling being British / American, I will do that.}

%\cdb{There are some inconsistencies in the bibliography. Sometimes we give DOIs, sometimes we don't. I would remove the DOIs everywhere, which would be easier than adding all the missing DOIs.} \cmb{I really like there being DOIs in references so I might just make sure they are added to every reference.} \cdb{That's fine by me, if you don't mind the extra work!}

%\cdb{Do we need to refer to Figure 1 somewhere in the text? Usually it's odd to have a figure without referring to it, but perhaps this isn't the case for visual abstracts?} \cmb{I would be happy not to refer to it.}

Understanding the deformation behaviour of polycrystalline materials is crucial for numerous industrial applications \cite{raabe2020current,roters2019damask}. These materials, composed of multiple grains with distinct crystallographic orientations, exhibit intricate microstructures that significantly influence their macroscopic mechanical properties. Moreover, the characterisation of localised deformations and microstructures formed during the deformation of polycrystalline materials is vital in developing a thorough physical understanding of the underlying mechanisms behind localisation phenomena such as local stress fields \cite{mianroodi2021teaching, khorrami2023artificial}, fracture and damage initiation \cite{roters2010overview, dao2001micromechanics, tasan2014strain, diehl2017coupled}, shear banding \cite{kasemer2020finite, jia2012non, kanjarla2010assessment,sedighiani2021large,STRSRD22}, and recrystallization nucleation \cite{kim2017mesoscopic, traka2021topological, shah2022coupling, chen2015integrated}.

Computational methods, particularly the finite element method (FEM) \cite{Quey2011, Sachtleber2002, Raabe2003, Bieler2009} and Fast Fourier Transform (FFT) \cite{roters2010overview,Sedighiani2020, Sedighiani2022, Eisenlohr2013}, have emerged as powerful tools for simulating the mechanical behaviour of virtual polycrystals \cite{Bulgarevich2023, STRSRD22, Shanthraj2015, Zhang2016}. The accuracy and the amount of detail that can be observed using these simulations strongly depends on the generated RVEs \cite{Diard2005, Lim2019, Ritz2008, STRSRD22}. A low-resolution simulation with simple cubic crystals is sufficient to predict macro-scale (global) data such as global crystallographic texture \cite{Sarma1996} or stress-strain response \cite{Diard2005}. However, a representative polycrystal morphology becomes essential to achieve a more detailed description of meso-scale deformation localisation effects \cite{STRSRD22, Diard2005, Ritz2008}. In addition, recent studies have emphasised the need for larger virtual polycrystals with more representative grain morphologies, considering inherent variability in both grain size and shape \cite{Bulgarevich2023, STRSRD22, Lim2019, Vermeij2024}. Therefore, constructing representative volume elements (RVEs) is essential in analysing the macroscopic and microscopic behaviour of polycrystalline materials \cite{Diard2005,Barbe2009}. Combining such representative computational microstructure with a proper materials model, such as the crystal plasticity model \cite{roters2010overview}, enables micro-scale analysis of many localised phenomena \cite{STRSRD22, shah2022coupling}.

Ensuring the accuracy of these analyses depends not only on having an appropriate constitutive law but also on careful reconstruction of the polycrystal's geometric features \cite{Vermeij2024, Liu2024}. Experimental efforts have contributed significantly to understanding real polycrystal morphologies, offering valuable insights into the size, morphology, and orientations of the crystals \cite{Sun2024, Pirgazi2014, Ghoncheh2020, Godec2020}. Nevertheless, creating representative microstructures with a large number of grains and authentic morphology still remains challenging.

One of the standard approaches to modelling polycrystalline materials in computational materials science is to represent them as a power diagram (also known as a Voronoi-Laguerre diagram) \cite{FWZL04,LLLFP11,QR18}, with each cell of the diagram corresponding to a distinct grain. Power diagrams, whose modern theory can be traced back to the 1980s \cite{IIM85,A87}, have found diverse applications, not just in microstructure modelling, but also in spatial analysis \cite{D08,SZYGSL15}, mesh generation \cite{GMMD14}, and in machine learning \cite{BCAB19}.  
Finding a power diagram with cells of prescribed volumes is in fact equivalent to solving an optimal transport problem where the target measure is a sum of Dirac masses. This goes back at least as far as \cite{Aurenhammer98}. For modern presentations in the computational geometry literature see \cite{qu2022power}, \cite[Chapter 6]{AurenhammerKleinLee}, in the optimal transport literature see \cite{KMT19}, \cite[Section 4]{MerigotThibertOT}, and in the microstructure modelling literature see \cite{BKRS20,BPR23}. This link ensures that power diagram-based approaches to modelling polycrystalline materials can generate large and complex microstructures in a matter of seconds, while requiring a relatively small number of parameters. 

A drawback of this approach is the idealised nature of the grains it produces
%A drawback of this approach is the {\db \sout{inherently}} {\db sometimes} \cmb{Flat boundaries are not inherently unrealistic? Perhaps change to "A drawback of this approach is the idealised nature of the grains it produces"?} unrealistic nature of the idealised grains it produces - 
 - they are convex and have flat boundaries. Moreover, any spatial anisotropy they possess is solely determined by the relative location of the seed points of neighbouring grains and not by the preferred growth directions of each grain or by the rolling direction during processing.
 %\cmb{That's a pretty long paragraph now, perhaps split?}
%\cdb{Done!}

An emerging approach to modelling polycrystals which addresses some of the limitations of power diagrams is to model them as anisotropic power diagrams (APDs) instead, as pioneered by \cite{ABGLP15}. In particular, APD-based modelling guarantees control over the anisotropy of individual grains and curved boundaries between neighbouring grains, with several such promising approaches explored in recent years by various authors \cite{altendorf20143d,ABGLP15,VBWBSJ16,SWKKCS18,TR18,PFWKS21,AFGK23,AFGK23B, jung2024analytical}. One obstacle to the wider adoption of APDs as a practical tool for modelling the microstructure of metals is the computational cost of generating them. Known optimal-transport-based efficient methods for generating power diagrams with grains of given volumes \cite{KMT19,MerigotThibertOT,BKRS20} do not translate to the anisotropic setup, and known techniques for generating APDs are drastically slower - while the usual runtime to generate a large power diagram with grains of given volumes is (tens of) seconds \cite{BKRS20, KSSB20}, for APDs it ranges from (tens of) minutes to (tens of) hours. This high computational cost associated with generating APDs poses a significant limitation to creating realistic microstructures with numerous grains and authentic morphology. This limitation is particularly pronounced in fields such as machine learning and data science, where generating a substantial number of representative microstructures is essential for a comprehensive study.

In this paper we develop a novel fast approach for generating APDs with prescribed statistical properties, in which we combine semi-discrete optimal transport techniques with modern GPU-oriented computational tools, originally developed for the Sinkhorn algorithm \cite{FPhD2020,FGCB20,CFGCD21}.  Using a single standard scientific-computing-oriented GPU, we achieve a three orders of magnitude speed-up versus a baseline CPU-only implementation, which ensures a near instantaneous computation of a generic large APD. As a result, we are able to: \begin{itemize}
    \item Fit an APD to a real EBSD measurement, provided by Tata Steel specifically for this publication, consisting of 4587 grains in 2D, with high variation in spatial anisotropy and grain volume, in \emph{under one minute}.
    \item Generate a realistic synthetic microstructure that is statistically equivalent to the EBSD measurement with 4587 grains in \emph{about four minutes}.
    \item Create an APD mimicking an EBSD scan of a bidirectionally 3D-printed stainless steel with long and thin grains in \emph{about 3 seconds}. 
\end{itemize} 

\section{Methods}
\subsection{Modelling polycrystalline materials with anisotropic power diagrams}
%\subsection{Preliminaries}
%\subsubsection{Notation and definitions} 
Let $\Omega \subset \R^D$ represent a 
%finite
bounded region occupied by a polycrystalline material. While our method applies to arbitrary nonconvex geometries, for simplicity we present it for the case when $\Omega$ is a 
%rectilinear
rectangular
domain (a rectangle if $D=2$ or a cuboid if $D=3$).
For $U \subset \Omega$, $|U|$ denotes the %area/volume of $U$. 
{area of $U$ if $D=2$ or the volume of $U$ if $D=3$.}

A matrix ${\mathbf A} \in \R^{D \times D}$ is symmetric positive definite if it satisfies 
\[
\mathbf A^{T} = \mathbf A \quad \text{and} \quad x \cdot \mathbf A x > 0 \; \; \forall \, x \in \R^D, \, {x \ne 0}.
\]
We refer to such matrices as \emph{anisotropy matrices} and denote the weighted norm they induce on $\R^D$ by $|\cdot|_{\mathbf A}$, that is $|x|_{\mathbf A} := \sqrt{x \cdot \mathbf A x}$.

Let $X = (x_i)_{i=1}^N \in \Omega^N$ be a set of distinct \emph{seed points} in $\Omega$, ${W = (w_i)_{i=1}^N \in \R^N}$ be a set of \emph{weights}, and $\Lambda = (\mathbf A_i)_{i=1}^N \in (\R^{D\times D})^N$ be a set of anisotropy matrices. The Anisotropic Power Diagram (APD) \cite{ABGLP15} given by the data $(X,W,\La)$
%\[
%{(X,W,\La) := \{(x_i,\mathbf{A}_i,w_i)\}_{i=1}^N}
%\]
%\cdb{The LHS does not equal the RHS. Might need to rething the notation slightly. Let's discuss it in the meeting on Thurs 18/01.}
is the tessellation $\{L_i\}_{i=1}^N$ of $\Omega$ defined by
\begin{equation}\label{def:L_i}
L_i := \Big\{x \in \Omega\, \mid\, |x-x_i|_{\mathbf A_i}^2 - w_i \leq |x-x_j|_{\mathbf A_j}^2 - w_j \; \; \forall \, j \in \{1,\dots,N\}\Big\}.
\end{equation}
Notable special cases of APDs occur when, for each $i$, we have 
{ (i) $\mathbf{A}_i = {\rm Id}$ and $w_i=0$, which results in a Voronoi diagram},
(ii) $\mathbf{A}_i = {\rm Id}$, which results in a power diagram, (iii) $w_i = 0$, which results in an anisotropic Voronoi diagram, (iv) $w_i = 0$ and $\mathbf{A}_i = c_i {\rm Id}$, $c_i \in \mathbb{R}$, which results in a Möbius diagram. The theory %behind (ii) and (iii) 
{of these diagrams is}
%are 
presented in \cite{BWY06}.

Let $V := (v_i)_{i=1}^N \in \R_+^N$ be a set of target volumes (if $D=3$) or target areas (if $D=2$), satisfying 
\[
v_i > 0, \qquad \sum_{i=1}^Nv_i = |\Omega|.
\]

In this paper we use the term \emph{single phase} to refer to APDs with grains of equal volume, i.e., $v_i = \frac{|\Omega|}{N}$ for each $i$, which represent idealised monodisperse microstructures, where are all the grains have essentially the same size. We use the term \emph{multi phase} to refer to polydisperse microstructures.

We call an APD generated by $(X,W,\La)$ \emph{optimal with respect to } $V$ if $|L_i| = v_i$ for {all $i \in \{1,\ldots,N\}$.} %$i=1,\dots,N$.

 We note that an anisotropy matrix $\mathbf A_i$ can carry volume information. Suppose we are given an APD in which each cell is non-empty. Fix $i \in \{1,\ldots,N\}$ and suppose we replace $\mathbf{A}_i$ by $c\mathbf{A}_i$ for some constant $c > 0$, while keeping $X$ and $W$ fixed, and keeping $\mathbf{A}_j$ fixed for all $j \ne i$.
 % It can be inferred from \eqref{def:L_i} that if all other anisotropy matrices and all points $X$ and all weights $W$ are kept fixed, replacing $\mathbf{A}_i$ by $c\mathbf{A}_i$ for some constant $c > 0$ will result 
% in an APD with volumes of $L_i$ and other cells changed. 
Simple calculations reveal that as we increase $c$, the volume of $L_i$ 
%shrinks
decreases
and, for $c$ large enough, the cell will have zero volume. Similarly, if $w_i > w_j$ for all $j\neq i$, by sending $c \to 0$, we find 
%ensure 
that $|L_i| \to |\Omega|$. At the same time, for any choice of the constant $c$, the ratio of eigenvalues of $c\mathbf{A}_i$ (and hence the target shape of $L_i$) remains unchanged. %{\db \sout{At the same time, for any choice of the constant $c$, the preferred anisotropy of $L_i$ remains unchanged.}}  \cdb{This is a but imprecise. The anisotropy of $L_i$ does change if it changes form $\emptyset$ to $\Omega$. I think you mean that the ratio of the eigenvalues of $\mathbf{A}_i$ does not depend on $c$. Perhaps it's easiest just not to mention this. Or you should say something like `At the same time, the ratio of the eigenvalues of $\mathbf{A}_i$ (and hence the target shape of $L_i$) remains unchanged.'}
To avoid such issues, we suggest normalising the anisotropy matrices so that ${\rm det} \, \mathbf A_i = 1$ for all $i$, which can be done while respecting the associated aspect ratios, as we will shortly explain.

To illustrate the geometric role that anisotropy matrices play, %$\mathbf A$ plays, 
we note that a two-dimensional anisotropy matrix $\mathbf A$ can be uniquely determined by three parameters $(a,b,\theta)$, where, in analogy with defining an ellipse, $a > 0$ is the major axis, 
%$b > 0$ 
{$ b \in (0,a]$}
is the minor axis, and 
%$\theta \in [0,\pi]$
$\theta \in {[0,\pi)}$
is the orientation angle. { To be precise,}
%The entries of %$\mathbf A$ are
%\[
%\mathbf A = \begin{bmatrix} a_{11} & a_{12} \\ a_{21} & a_{22} \end{bmatrix},\quad \begin{cases} a_{11} &=a^{-2}\cos^2(\theta) + b^{-2}\sin^2(\theta), \\ a_{12} &= \, a_{21}\,  = (a^{-2} - b^{-2})\cos(\theta)\sin(\theta), \\ a_{22} &=a^{-2}\sin^2(\theta) + b^{-2}\cos^2(\theta). \end{cases}
%\]
\begin{equation}
\label{eq: A}    
{
\mathbf A{(a,b,\theta)} = %\mathbf A %(a,b,\theta) = 
\mathbf{V} \mathbf{D}
\mathbf{V}^{-1},
\quad \textrm{where}
\quad
\mathbf{V} = 
\begin{pmatrix}
\cos \theta & -\sin \theta 
\\ 
\sin \theta & \cos \theta
\end{pmatrix}, 
\quad
\mathbf D=
\begin{pmatrix}
a^{-2} & 0 
\\
0 & b^{-2}
\end{pmatrix}.
}
\end{equation}
Note that $|x|_{\bm A} = 1$ is the equation of the ellipse with major axis $a$, minor axis $b$ and orientation angle $\theta$.

Given an 
%unnormalised 
{anisotropy matrix}
$\mathbf A(a,b,\theta)$, its normalised counterpart $\hat{\mathbf A}(\hat a, \hat b, \theta)$, satisfying ${\rm det} \, \hat{\mathbf A} = 1$, 
%in which the anisotropy ratio $\frac{a}{b}$ is preserved, 
is determined by $(\hat a, \hat b, \theta)$, where 
%$\hat a = \sqrt{\frac{a}{b}},\quad \hat b = \frac{1}{\hat a}$. 
$\hat a = \sqrt{a/b}$, $\hat b=1/\hat a$.
The anisotropy ratio $a/b$ is preserved by the normalisation: $\hat a/\hat b= a/b$.
A similar comment applies in 3D with ellipsoids, which are generated by six parameters $(a,b,c,\theta,\phi,\gamma)$, the major, middle and minor axes and the Euler angles.

\subsection{Finding optimal APDs} 
Given a set of seed points $X = (x_1, \dots, x_N)$, a set of anisotropy matrices $\La = (\mathbf A_1, \dots, \mathbf A_N)$, and a set of target volumes $V= (v_1, \dots, v_N)$, the problem of finding weights $W$ such that the APD generated by $(X,W,\La)$ is optimal with respect to $V$, i.e., that $|L_i| = v_i$ for all $i$, is a \emph{semi-discrete optimal transport problem}; see for example \cite[Section 4]{MerigotThibertOT} or \cite[Chapter~5]{PC19}. In particular, it can be solved by maximising the continuously differentiable, concave function  
\begin{equation}\label{eqn-g}
g(W) := \sum_{i=1}^N \left( (v_i - |L_i|)w_i + \int_{L_i} |x-x_i|_{\mathbf A_i}^2\,dx \right),    
\end{equation}
where the cells $L_i$ are computed from the weights $w_1, \dots, w_N$ as in~\eqref{def:L_i}. We observe that $g$ is concave and 
its gradient is
\begin{equation}\label{eqn-grad-g}
\left(\nabla g(W)\right)_i = v_i - |L_i|.
\end{equation}
%and thus if $W$ maximizes $g$, then clearly an APD generated by $(X,W,\La)$ is optimal with respect to $V$. 
Thus $W$ maximises $g$ if and only if the APD generated by $(X,W,\La)$ is optimal with respect to $V$.
%\cmb{describe properties, a lemma about dual functional (concave, C1, gradient formula). Discuss that if all matrices are transformed in the same manner from identity, we still have straight lines as edges}
This is well-known in the optimal transport literature (see for example \cite{MerigotThibertOT} and \cite{PC19}). In the isotropic case, when $\mathbf{A}_i = {\rm Id}$ for all $i$, this was first applied in the context of microstructure modelling in  \cite{BKRS20} and then subsequently in papers including \cite{KSSB20} and \cite{BPR23}.

In practice, $\nabla g(W) = 0$ is solved up to relative error tolerance 
\begin{equation}
\label{eq: epsilon}
\frac{|(\nabla g(W))_i|}{v_i} = \frac{\big||L_i|-v_i\big|}{v_i} \leq \varepsilon,
\end{equation}
where, e.g., setting $\varepsilon = 0.01$ corresponds to allowing grain size deviation of up to 1\%.

Similarly, the integral in \eqref{eqn-g} and the area/volume $|L_i|$ in \eqref{eqn-g} and \eqref{eqn-grad-g} are in practice approximated with sums over a discretisation of the domain $\Omega$ by pixels/voxels. %\cmb{This paragraph might go since we introduce $M$ in the next section.}

We will now describe a fast implementation of algorithms for computing APDs via \eqref{def:L_i} and for finding optimal APDs with respect to prescribed volumes, namely finding 
\[
W \in {\rm argmax}\, g.
\]

% In practice, $\nabla g(W) = 0$ is solved up to relative error tolerance 
% \begin{equation}
% \label{eq: epsilon}
% \frac{|(\nabla g(W))_i|}{v_i} = \frac{\big||L_i|-v_i\big|}{v_i} \leq \varepsilon,
% \end{equation}
% where, e.g., setting $\varepsilon = 0.01$ corresponds to allowing grain size deviation of up to 1\%.

% Similarly, the integral in \eqref{eqn-g} and the area/volume $|L_i|$ in \eqref{eqn-g} and \eqref{eqn-grad-g} are in practice approximated with sums over a discretisation of the domain $\Omega$ with $M^D$ pixels/voxels (i.e., $M$ pixels/voxels in each spatial dimension). We refer to $M$ as the inverse pixel length parameter, and $D$ is the dimension of the problem.

\subsection{Implementation}\label{sec:pixel_method}

%\cmb{removed definition environment}
%\begin{definition}[Pixel-discretised domain]\label{def:discretised_domain}
Given a domain $\Omega \subset \R^D$, 
we let $P(y,s)$ denote the
%consider a
pixel/voxel centred at $y \in \Omega$ 
%and with a length vector 
with side lengths
$s \in \R_+^D$, %containing the side lengths of the pixel,
that is,
\[
P(y,s) := \left\{ y + {\rm diag}(s) t \,\mid\, t \in \left(-\tfrac12,\tfrac12\right)^D \right\},
\]
where $\mathrm{diag}(s)$ denotes the $D$-by-$D$ diagonal matrix with $s$ on the diagonal.
%which implies that 
The area/volume of the pixel $P(y,s)$ is $|P(y,s)| = { s_1 s_2  \cdots s_D = \rm det}\left({\rm diag}(s)\right)$. 
We call a collection of pixels/voxels generated by $(Y,s) := \{(y_j,s)\}_{j=1}^{J}$, $J \in \N$, a \emph{discretisation} of $\Omega$, if 
\[
P(y_j,s) \cap P(y_k,s) = \emptyset\, \text{ for all }\,j \neq k,\, \text{ and 
}\, 
%\left| 
\Bigg|
|\Omega| - \sum_{j=1}^{J} |P(y_j,s)|
%\right| 
\Bigg|
\leq 
%\varepsilon
\delta
,
\]
where 
%$\varepsilon \geq 0$ 
$\delta \geq 0$
is a tolerance parameter. 
%\cdb{We use $\varepsilon$ for the volume tolerance.}
%The square 
Square pixels/cubic voxels are obtained by setting 
\begin{equation}\label{regular-grid}
    s = \frac{1}{M} \mathbf{1}_D, \quad \mathbf{1}_D = (1,\dots,1) \in \R^D,
\end{equation}
where $M \in \N_+$ is the resolution parameter. In particular, if ${\Omega = [0,1]^D}$ and $\varepsilon = 0$, the discretised domain is then the regular grid of $J = M^D$ pixels/voxels, where $|P| := |P(y_j,s)| = J^{-1} =  M^{-D}$ is the area/volume of the pixel/voxel.
%\end{definition}

Algorithm~\ref{alg:pixel_APD} is a standard method for numerically computing APDs. The novelty of our work is an efficient GPU implementation of this algorithm, as described below in Section \ref{subsec:GPU}.

%\begin{example}[Uniform discretisation]\label{ex:uniform_discretisation}
%    A uniform discretisation is obtained by setting 
%    \[
%    s = \frac{1}{M} \mathbf{1}_D, \quad \mathbf{1}_D = (1,\dots,1) \in \R^D.
%    \]
%    where $M \in \R_+$ is the inverse pixel/voxel length parameter. In particular, if ${\Omega = [0,1]^D}$ and $\varepsilon = 0$, the discretised domain in the sense of Definition~\ref{def:discretised_domain} is then the regular grid of $J = M^D$ pixels/voxels, where $P = J^{-1} =  M^{-D}$.
%\end{example}

\begin{algorithm}[H]
   \caption{Pixel method for computing an APD}\label{alg:pixel_APD}
    \begin{flushleft}
    \textbf{Input:} $D \in \{2,3\}$ (the dimension), $\Omega \subset \R^D$ (the domain), $N \in \N$ (the number of grains), 
    %$X = \{x_i\}_{i=1}^N \subset \Omega$ 
    $X = (x_i)_{i=1}^N \in \Omega^N$
    (seed points), %$\La = \{\mathbf A_i\}_{i=1}^N$
    $\La = (\mathbf A_i)_{i=1}^N$
    (anisotropy matrices), 
    %$W = \{w_i\}_{i=1}^N$
    $W = (w_i)_{i=1}^N$
    (weights), $J \in \N$ (number of pixels/voxels), $(Y,s) \in (\R^D)^J \times (\R^D)$ (discretised domain / collection of pixels/voxels).\\
    \textbf{Output:} An assignment vector $H \in \N^{J}$, where the $j$th pixel is assigned to the grain $H_j \in \{1, \dots, N\}$.
    \\
    \textbf{Procedure:}
    \end{flushleft}
   \begin{algorithmic}[1]
   \State Compute the cost matrix $\mathbf C \in \R^{N \times J}$, where $\mathbf C_{ij} := |y_j - x_i|^2_{A_i} - w_i$.
   \State Set $H_j \in \argmin_{i \in \{1,\dots N\}}\mathbf C_{ij}$.
\end{algorithmic}
\end{algorithm}

Next we describe how optimal APDs with cells of given areas/volumes can be generated by combining semi-discrete optimal transport theory with the pixel method. Consider
%Given 
a domain $\Omega \subset \R^D$ and its discretisation $(Y,s) = \{(y_j,s)\}_{j=1}^J$, as well as the set of seed points 
%$X =\{x_i\}$, 
$X =(x_i)_{i=1}^N$,
%and 
a set of anisotropy matrices
$\La = (\mathbf A_i)_{i=1}^N$,
%$\La = \{\mathbf A_i\}$, 
and a set of target areas/volumes
$V =(v_i)_{i=1}^N$.
Using the regular discretisation from \eqref{regular-grid}, we approximate %the pixel method concerns approximating 
the dual objective functional $g$ from \eqref{eqn-g} with the discretised function $\tilde{g}:\mathbb{R}^N \to \mathbb{R}$ defined by 
\begin{equation}\label{eqn-approx-g}
\tilde g(W) := \sum_{i=1}^N 
\left( 
\Bigg(
%\left(
v_i - \sum_{\substack{j \in \{1,\dots,J\}\\H_j = i}}|P|
%\right)
\Bigg)
w_i + \sum_{\substack{j \in \{1,\dots,J\}\\H_j = i}}|P| \, 
%|y_j-x_i|_{A_i}^2
|y_j-x_i|^2_{\mathbf{A}_i}
\right),
\end{equation}
%\cdb{The big \textbackslash{}left( and \textbackslash{}right) brackets look a bit ugly (way too big), but I don't know what we can do about them. We could split up the sum over $i$ into two sums of $i$, to avoid the need for the outer brackets? Or we could just leave it as it is. I'm happy either way.}
where $H_j$ was defined in Algorithm~\ref{alg:pixel_APD}. 
%Note that 
The gradient of $g$ can be approximated by
\begin{equation}
\label{eqn-approx-grad-g}
%\left(\nabla\tilde g(W)\right)_i = v_i - \sum_{\substack{j \in \{1,\dots,J\}\\H_j = i}}|P|.
\left(\nabla g(W)\right)_i \approx 
v_i - \sum_{\substack{j \in \{1,\dots,J\}\\H_j = i}}|P|.
\end{equation}
%\cdb{I replaced $\tilde{g}$ by $g$ and $=$ by $\approx$ since $\tilde{g}$ is not differentiable.}
Note that this is not precisely the gradient of $\tilde{g}$, which is not differentiable everywhere since  $H_j$ is a piecewise constant function of $W$.

We can find an optimal APD with grains of areas/volumes $v_i$ by maximising the discretised objective function $\tilde{g}$, as described in Algorithm~\ref{alg:optimal_APD}. In Section \ref{subsec:GPU} we describe an efficient GPU implementation of this algorithm.
%The precise algorithm is as follows. 

\begin{algorithm}[H]
   \caption{Generating an optimal APD using optimal transport theory}\label{alg:optimal_APD}
    \begin{flushleft}
    \textbf{Input:} $D \in \{2,3\}$ (the dimension), $\Omega \subset \R^D$ (the domain), $N \in \N$ (the number of grains), 
    %$X = \{x_i\}_{i=1}^N \subset \Omega$ 
    $X = (x_i)_{i=1}^N \in \Omega^N$
    (seed points), %$\La = \{\mathbf A_i\}_{i=1}^N$
    $\La = (\mathbf A_i)_{i=1}^N$
    (anisotropy matrices),
    $J \in \N$ (initial number of pixels/voxels), 
    %$J \in \N$ (discretisation parameter), 
    %$V = \{v_i\}_{i=}^N$
    $V = (v_i)_{i=}^N$
    (target volumes), %$\varepsilon \in \R_+$ 
    $\varepsilon > 0$
    (relative tolerance).\\
    \textbf{Output:} The generators %$(X,\La,W)$
    $(X,W,\La)$
    of an APD optimal with respect to $V$, up to a relative error tolerance $\big||L_i|-v_i\big| \leq \varepsilon v_i $.\\
    \textbf{Procedure:}
    \end{flushleft}
   \begin{algorithmic}[1]
   \State Generate a discretisation $(Y,s)$ of $\Omega$ with $J$ pixels/voxels. %given by $(Y,s)$.
   \State Set the initial guess $W_0$ to $W_0 = (0,\dots,0) \in \R^N$. 
   \State Use  a  numerical  optimisation  method, such as the classic L-BFGS method \cite{LN89}, to find $W$ that maximises the function $\tilde g$ defined in \eqref{eqn-approx-g}, starting from the initial guess $W_0$. Terminate when 
   %the $\nabla \tilde g$, defined in \eqref{eqn-approx-grad-g} satisfies $|(\nabla\tilde g(W))_i| \leq \varepsilon v_i$ 
   $|v_i - m_i |P| \,|  < \varepsilon v_i$
   %\cdb{$\nabla \tilde{g} is not defined.$}
   for all $i \in \{1,\dots,N\}$, where $m_i$ is the number of pixels/voxels in cell $i$,  $m_i = \# \{ j \in \{1,\ldots,J\} \, | \,  H_j = i \}$.
   \State If it is not possible to hit the desired tolerance $\varepsilon$, then increase $J$ %refine your discretisation 
   and restart.  
\end{algorithmic}
\end{algorithm}

For simplicity, we have presented the algorithms for discretisations of $\Omega$ by regular rectangular grids. In principle, however,
%Note that the method is not restricted to uniform discretisations of $\Omega$
%and in principle 
any tessellation of $\Omega$ could 
%can
be used. 
%In particular, 
For example,
given a triangulation of $\Omega \subset \mathbb{R}^2$ by $J$ triangles, 
the corresponding 
objective function $\tilde{g}$ in Algorithm~\ref{alg:optimal_APD} is
\[
\tilde g(W) := \sum_{i=1}^N 
\left( 
\Bigg(
%\left(
v_i - \sum_{\substack{j \in \{1,\dots,J\}\\H_j = i}}|P_j|
%\right)
\Bigg)
w_i + \sum_{\substack{j \in \{1,\dots,J\}\\H_j = i}}|P_j| \, 
%|y_j-x_i|_{A_i}^2
|y_j-x_i|^2_{\mathbf{A}_i}
\right),
\]
where $y_j$ is the centroid of triangle $j$ and $|P_j|$ is its area.
%the centroid point of the $i$th triangle 
% can be taken as the point $y_i$, and $|P_j|$ would be the area of the triangle, replacing the current uniform area/volume $|P|$. 
Using such Finite Element Method-friendly triangulations might prove useful when using our method side by side with crystal plasticity simulations.

To generate realistic artificial microstructures, we will also employ a generalised version of Lloyd's algorithm \cite{DFG99}, which was introduced for the isotropic case ${\mathbf{A}_i= {\rm Id}}$ for all $i$ in \cite[Algorithm 2]{BKRS20} in the setting of microstructure modelling. The purpose of the generalised Lloyd's algorithm, stated in Algorithm~\ref{alg:Lloyds}, is to generate more realistic, `regular' microstuctures, where
%, {\db the seeds tend to be more uniformly distributed (compared to sampling them at random)} \cmb{only when masses are uniformly distributed? I am not sure we want to mention that? It sounds restrictive, as if it always has to happen and it is a limitation of the method, which isn't really the case right? I would just focus on the simple-connectedness} and where 
the cells tend to be simply-connected, which need not be the case for APDs in general (APD cells can be disconnected and have holes). Unlike Algorithm~\ref{alg:optimal_APD}, Algorithm~\ref{alg:Lloyds} does not require seeds $X$ as an input.

\begin{algorithm}[H]
\caption{Generalised Lloyd's algorithm}\label{alg:Lloyds}
    \begin{flushleft}
    \textbf{Input:} $D \in \{2,3\}$ (the dimension), $\Omega \subset \R^D$ (the domain), $N \in \N$ (the number of grains), 
    %$\La = \{\mathbf A_i\}_{i=1}^N$
    $\La = (\mathbf A_i)_{i=1}^N$
    (anisotropy matrices),
    $J \in \N$   (initial number of pixels/voxels), 
    %{\db \sout{$W = \{w_i\}_{i=1}^N$ (weights),}} 
    %{\db \sout{$P \in \N$ (discretisation parameter),}} 
    %$V = \{v_i\}_{i=}^N$
    $V = (v_i)_{i=}^N$
    (target volumes), 
    %$\varepsilon \in \R_+$ 
    $\varepsilon > 0$
    (relative tolerance) and $K \in \N$ (number of regularisation steps).\\
    \textbf{Output:} The generators %$(X,A,W)$ 
    $(X,W,\Lambda)$
    of an APD optimal with respect to $V$, up to a  relative error tolerance $\big||L_i|-v_i\big| \leq \varepsilon v_i $.
    \end{flushleft}
\begin{algorithmic}[1]
    %\State Generate a discretised domain of $\Omega$ given by $(Y,S)$.
    \State Pick or randomly select $N$ initial seed points 
    %$X^{0} = \{x^0_i\}_{i=1}^N \subset \Omega$.
    $X^{0} = (x^0_i)_{i=1}^N \in \Omega^N$.
    \State Set the initial guess $W^0_0$ to $W^0_0 = (0,\dots,0) \in \R^N$.
    %\For{$k = 0,1,\dots, K$}
    \For{$k = 0,1,\dots, K-1$}
        \State Use Algorithm~\ref{alg:optimal_APD} with initial guess $W_0^k$ to find an APD generated by $(X^k,W^k,\La)$
        that
        is optimal with respect to $V$.
        \State Using the discretisation $(Y,S)$ generated 
        % in
        by Algorithm~\ref{alg:optimal_APD}, set $x_i^{k+1}$ to be the discrete centroid of the $i$th grain:
        \[x^{k+1}_i = \frac{1}{\sum_{\substack{j \in \{1,\dots,J\}\\H_j = i}}|P|}\sum_{\substack{j \in \{1,\dots,J\}\\H_j = i}}|P|y_j,\]
        thus obtaining $X^{k+1}$.
        %\csr{Is $P$ a constant for all cells? If not then does it need to be evaluated at $s_j$? See earlier comment about $P$ the volume and $P$ the cell. Or is it kept here to make it obvious that it is the centre of mass that we're calculating.}
        \State Set $W^{k+1}_0 = W^{k}$.
        \EndFor
        \State Use Algorithm~\ref{alg:optimal_APD} with initial guess $W_0^K$ to find an APD generated by $(X^K,W^K,\La)$
        that is optimal with respect to $V$. Set $X = X^K$, $W = W^K$.
\end{algorithmic}
\end{algorithm}

\subsubsection{Choosing the discretisation resolution}\label{sec-choose-M}
%\cmb{Move all the discussion about choosing pixels here}
For a discretisation with square pixels/cubic voxels of 
%size
area/volume
$|P| = |\Omega| M^{-D}$, to 
reach
the desired tolerance $\varepsilon >0$ in Algorithm~\ref{alg:optimal_APD}, 
it is usually necessary
%{\db \sout{one requires}}
that $|P| < \varepsilon v_i$ for all $i=1, \dots, N$  since this is the relative error of misassigning a single pixel/voxel to cell $i$. For example, if $v_i = \frac{|\Omega|}{N}$ for all $i$ (single-phase material), 
%a 
the
theoretical lower bound $|P| < \varepsilon v_i$ gives
%on the choice of $M$ is thus
\begin{equation}\label{eqn-lower-bdd-M}
\frac{|\Omega|}{M^D} < \frac{\varepsilon |\Omega|}{N} \quad \iff \quad M > 
%\left(\frac{N}{\varepsilon |\Omega|}\right)^{1/D}.
\left(\frac{N}{\varepsilon }\right)^{1/D}.
\end{equation}
In Table \ref{table:choose_M} we display the values of $M$ given by the lower bound \eqref{eqn-lower-bdd-M} for various values of $N$ and $D$ for $\varepsilon = 0.01$.

\begin{table}[h!]
\begin{center}
\begin{tabular}{|l||c|c|c|c|c|c|c|c|c|}
\hline
\diagbox{$D$}{$M$}{$N$} & 25 & 50 & 100 & 250 & 500 & 1000 & 2500 & 5000 & 10000 \\
\hline
2 & 50 & 71 & 100 & 158 & 224 & 316 & 500 & 707 & 1000 \\
3 & 14 & 17 & 22 & 29 & 37 & 46 & 63 & 79 & 100 \\
\hline
\end{tabular}
\end{center}
\caption{The theoretical lower bound \eqref{eqn-lower-bdd-M} on $M$,  the pixel sampling resolution along the axes of the domain $\Omega = [0,1]^D$, when targeting a relative volume accuracy of 
$\varepsilon = 0.01 = 1\%$ (in a single phase material), for different choices of the number
of seed points $N$ and the dimension $D$ of the domain. In practice, for most problems, choosing $M$ equal to two times the value
given by the lower bound \eqref{eqn-lower-bdd-M}
is sufficient (see equation \eqref{choice-of-M}).}
%\cdb{I think there is a discrepancy between this comment and equation \eqref{choice-of-M}. In equation \eqref{choice-of-M} we halve the value of $|P|$, which is not the same as doubling the value of $M$. Which method do we use?}}
\label{table:choose_M}
\end{table}

This may not always be sufficient, however, since changing the weights of an APD typically reassigns several pixels at the same time, and so in practice the resolution parameter $M$ is always choosen to be the smallest integer such that 
\begin{equation}\label{choice-of-M}
%{\mb |\Omega|\left(\frac{2}{M}\right)^D < \varepsilon \min_i v_i}
\frac{{|\Omega|}}{M^D} < \frac{\varepsilon}{2^D} \min_{i} v_i,
\end{equation}

%{\mb \sout{take} recommend taking}

%The resolution parameter $M$ from \eqref{regular-grid} has to be chosen large enough to ensure that we can achieve the desired relative error tolerance determined by $\varepsilon$.

%We discretise $\Omega$ with $M^D$ square pixels/cubic voxels. {\mb As was discussed in Section~\ref{sec:pixel_method}, t}he parameter $M$ has to be chosen large enough to ensure that we can achieve the desired relative error tolerance $\varepsilon${\mb . \sout{, see Section~\ref{sec:pixel_method}}}. Specifically, we always choose the smallest integer $M$ such that
% \begin{equation}\label{choice-of-M}
% %{\mb |\Omega|\left(\frac{2}{M}\right)^D < \varepsilon \min_i v_i}
% \frac{{|\Omega|}}{M^D} < \frac{\varepsilon}{2^D} \min_{i} v_i,
% \end{equation}

\subsubsection{GPU acceleration and kernel operations} 
\label{subsec:GPU}

The theoretical cost of Algorithm~\ref{alg:pixel_APD} in Section~\ref{sec:pixel_method} is $\mathcal{O}(NJ)$, where $N$ is the number of seed points and $J$ is the number of pixels/voxels that we use to discretise the domain. For a discretisation by regular rectangular grids given by \eqref{regular-grid}, $J = M^D$ and if $M$ satisfies \eqref{choice-of-M}, then we obtain the quadratic scaling $\mathcal{O}(NJ) = \mathcal{O}(N^2)$ for Algorithm \ref{alg:pixel_APD} (and Algorithm \ref{alg:optimal_APD} scales at least quadratically), with typically a very large prefactor. As a result, for typical values of $N$ of interest, a standard implementation of such algorithms will result in runtimes ranging from minutes to hours.

In order to prevent this computation from becoming a numerical bottleneck, we turn to GPU computing. In particular, we employ the GPU acceleration architecture provided by the machine-learning library \textsc{PyTorch} \cite{PyTorch} and rely on its in-house L-BFGS solver for Algorithm~\ref{alg:optimal_APD}, packaged as a general minimisation tool via the \textsc{PyTorch Minimize} library \cite{F21}. Notably,  we rely on the very fast automatic differentiation available in \textsc{PyTorch} to quickly compute machine-precision accurate derivatives of $\tilde g$ from \eqref{eqn-approx-g}. This works remarkably well and is in fact quicker than providing the gradients by hand using the formula in \eqref{eqn-approx-grad-g}, even though $\tilde g$ is not everywhere differentiable -- this is, however, in agreement with recent literature on this topic \cite{LYRY20}.

To avoid memory overflows when assembling the cost matrix $\mathbf C \in \R^{N \times J}$ in Algorithm~\ref{alg:pixel_APD}, we employ the kernel operations library  \textsc{PyKeOps}, an extension for \textsc{PyTorch} that provides efficient support for distance-like matrices \cite{CFGCD21}. As detailed in \cite{FGCB20,feydy2019interpolating,FPhD2020}, turning to a \textsc{PyKeOps} backend brings the memory footprint of optimal transport solvers from $\mathcal{O}(NJ)$ to  $\mathcal{O}(N+J)$ and provides a $\times 10$ to $\times 100$ speed-up versus baseline \textsc{PyTorch} implementations.

As will be presented in various examples, this cut down the typical runtime of our method to (tens of) seconds, making it a feasible tool for generating large samples of realistic random volume elements. We publish our code as a Python repository \textsc{PyAPD} \cite{PyAPD}.

\section{Numerical results}
% The central result of our paper is a fast implementation of algorithms for computing APDs and finding optimal APDs with respect to prescribed volumes, namely finding 
% \[
% W \in {\rm argmax}\, g.
% \]
Due to its speed, our implementation of the algorithms allows us to tackle several applications that, up to now, would have been considered prohibitively expensive. In what follows, we present a comprehensive list of examples showcasing the speed and the versatility of our method. With regards to speed, in Section~\ref{sec:runtime} we present runtime tests for computing APDs for given inputs $(X,\La,W)$, as well as for generating optimal APDs with cells of prescribed volumes. This is followed by examples based on Electron Backscatter Diffraction (EBSD) measurements provided by Tata Steel. 
First, in Section~\ref{sec:ex_ebsd}, we fit optimal APDs to the EBSD data. Then, in Section~\ref{sec:ebsd-synthetic}, we demonstrate how to generate realistic, synthetic microstructures by sampling from a joint probability distribution of grain volumes, aspect ratios and orientations, which is obtained as a fitted kernel density estimator \cite{S18} of the EBSD data.
%, followed by finding optimal APDs for generated samples of artificial representative volume elements, with aspect ratios, orientations and volumes drawn from a joint probability distribution, obtained as a maximum likelihood fit of the EBSD data. 
Finally, in Section~\ref{sec:ex_3d_printed_steel}, we give an example of how to generate a complex microstructure representing a 3D-printed stainless steel. %{\db \sout{Finally, to further showcase the versatility of the approach, we discuss how, with minimal effort, we are able to generate microstructure of desired geometry.}} \cdb{Possibly overselling it a bit?} \cmb{yes :) By the way, our fit is no longer a Maximum Likelihood fit, but rather we use a kernel density estimator to approximate the probability distribution.}

We refer to our Python repository, \textsc{PyAPD} \cite{PyAPD}, where readers can find Jupyter notebooks detailing each of the examples presented. 

\subsection{Hardware}\label{sec:runtime-hardware}
The speed of the method relies heavily on the GPU at our disposal. The relevant baseline against which GPUs should be compared are floating-point operations per second (FLOPS). See Table~\ref{table:FLOPS}.
\begin{table}[h!]
\begin{center}
\begin{tabular}{|c || c | c |} 
 \hline
 \bfseries GPU type & \bfseries Float32 FLOPS & \bfseries Float64 FLOPS \\ [1.5ex] 
 \hline\hline
 NVIDIA A100 & 
19.49 TFLOPS & \bfseries 9.746 TFLOPS  \\[1ex]
 \hline
 NVIDIA Tesla T4 & 8.141 TFLOPS & 0.254 TFLOPS  \\[1ex]
 \hline
 \hline
 NVIDIA GeForce RTX 4090 & \bfseries 82.58 TFLOPS & 1.29 TFLOPS \\[1ex]
 \hline
 AMD Radeon RX 7600 & 21.75 TFLOPS & 0.679 TFLOPS \\[1ex]
 \hline
\end{tabular}
\end{center}
\caption{Comparison of FLOPS performance of various popular GPUs. A100 and Tesla T4 are professional GPUs geared towards scientific computing and machine learning, whereas the other two are commercially available and video-gaming oriented. As of October 2023, GeForce RTX 4090 is the most powerful gaming-oriented GPU available and AMD Radeon RX 7600 is a popular affordable mid-range GPU. Note the drastic decrease in performance between single precision (Float32) and double precision (Float64) arithmetic (between $ 30\times$ and $65\times$ decrease) in all cases except for A100 ($\sim 2\times$ decrease). Note that 1 TFLOP denotes $10^{12}$ (1 trillion) floating point operations per second.}
\label{table:FLOPS}
\end{table}

We perform our numerical experiments on a single A100 GPU, available through the NERSC high-performance computing cluster Perlmutter (see Acknowledgements), but readers are invited to test it for themselves using a T4 Tesla GPU offered free of charge by Google Colab \cite{GoogleColab} in a notebook we provide, see \cite{PyAPD}.

\subsection{Runtime tests}\label{sec:runtime}
%\paragraph{\textbf{Setup}}\label{sec:runtime-setup}
We will present the following two sets of runtime tests.
\begin{enumerate}[(a)]
    \item Use of Algorithm~\ref{alg:pixel_APD} to compute an  APD, i.e., the tessellation $\{L_i\}_{i=1}^N$ from \eqref{def:L_i}, for a fixed triple $(X,W,\La)$.
    \item Use of Algorithm~\ref{alg:optimal_APD} to find optimal APDs with cells of prescribed volumes, %for artificially generated 
    to generate artificial single-
    and multi-phase 
    %collection of grains 
    microstructures
    in 2D and 3D. 
\end{enumerate}

The common setup is as follows. We consider the box $\Omega = [0,1]^D$, $D=2,3$, with $N$ grains.
%, where, if 
If
$D=2$, we take
\begin{equation}\label{eqn-N_choice_2D}
N \in \{25, 50, 100, 250, 500, 1000, 2500, 5000\}.
\end{equation}
If $D=3$, we take
\begin{equation}\label{eqn-N_choice_3D}
N \in \{50, 100, 250, 500, 1000, 2500, 5000,10000\}.
\end{equation}

In each case the seed points $X = \{x_i\}_{i=1}^N$ are drawn randomly from the uniform distribution on  $[0,1]^D$, but, in the runtime test (b), to increase numerical stability, the sampling is sequential and any new sampled point $x \in \Omega$ is accepted only if it is not too close to some previously sampled seed point $x_i$, namely, if  $|x-x_i| > C N^{-1/D}$, where $C=\tfrac{2}{10}$. This is motivated by the fact that on a regular grid containing $N$ points, the seeds would be distance
%they 
%would be 
$N^{-1/D}$ apart. Setting $C=\tfrac{2}{10}$ ensures that it is a very mild constraint and in fact only a small fraction of the sampled points get rejected. On average, in 2D about 7\% of proposed seed points get rejected, whereas in 3D it goes down to only about 2\%. At the same time, we avoid situations where two seed points are almost exactly on top of each other. In our tests, we saw that this issue 
%leading 
led
to unusually long runtimes for some random runs, especially for large multi-phase problems with small anisotropy. Note that the sampling is done using the random number generator from the machine-learning library \textsc{PyTorch} and, even with the exclusion, it is almost instantaneous.

We sample normalised anisotropy matrices $\hat{\mathbf{A}}_i$
%so that
satisfying the constraint
$\rm{det} \, \hat{\mathbf A}_i = 1$ for all $i$. In 2D this is achieved by 
sampling
\[
\hat a \sim {\rm Uniform}(1-\alpha,1), \quad  \theta \sim {\rm Uniform}(0,\pi),
\] 
where
$\alpha \in [0,1)$ is the anisotropy threshold parameter. Then we define a normalised anisotropy matrix $\hat{\mathbf A}=\hat{\mathbf A}(1/\hat a,\hat a,\theta)$, as descried above, 
where $1/\hat a$ and $\hat a$ are the major and minor axes, and $\theta$ is the rotation angle.
%assembling them from random collections of $\{(\hat a, %\tfrac{1}{\hat a}, 
%{\db 1/\hat a},
%\theta)\}$, where 
%\[
%\hat a \sim {\rm Uniform}(1-\alpha,1), \quad  \theta \sim {\rm Uniform}(0,2\pi),
%\] 
%where $\hat a$ and $\tfrac{1}{\hat a}$ are the axes, $\theta$ is the rotation angle and   $\alpha \in [0,1]$ is the anisotropy threshold parameter $\alpha \in [0,1]$. 
Note that setting $\alpha = 0$ corresponds to  the fully isotropic case ($A_i = {\rm Id}$ for all $i$), whereas setting $\alpha$ close to 1 means we accept any level of anisotropy. 

Similarly, in 3D the determinant constraint is achieved by assembling the matrices from collections $(\hat a, \hat b, 1/(\hat a \hat b), \theta, \phi,\gamma)$, where
\[
\hat{a} \sim {\rm Uniform}(1-\alpha,1), \quad \hat{b} \sim {\rm Uniform}\left(1-\alpha, \frac{1}{1-\alpha}\right),\quad  \theta,\phi,\gamma \sim {\rm Uniform}(0,2\pi).
\]
Here $\hat a, \hat b, 1/(\hat a \hat b)$ are the axes and $\theta,\phi,\gamma$ are rotation angles and again $\alpha \in [0,1)$ is the anisotropy threshold parameter, with $\alpha=0$ again corresponding to the isotropic case. 
%\cmb{this is discussed few lines above for 2D.}
%\cdb{Should $\theta,\phi,\gamma$ be sampled from different ranges?} \cmb{Potentially if we want to do it in a "minimal" way, but isn't it also true that if we do it in a "non-minimal" way and sample them all uniformly from $(0,2\pi)$, then all the duplicates appear the same amount of times so we retain uniformity? Similarly to the 2D case where sampling from $(0,2\pi)$ leads to each potential configuration corresponding to two choices of $\theta$.}

In both the 2D and 3D examples, we always report runtimes for three choices of the anisotropy threshold parameter, namely $\alpha = 0,\,0.3,\,0.7$.

In runtime test (b) we set the relative error tolerance of the areas/volumes of the grains to be 1\%, meaning that $\varepsilon = 0.01$ (see equation \eqref{eq: epsilon}). We discretise $\Omega$ with $M^D$ square pixels/cubic voxels (see \eqref{regular-grid}). The resolution parameter $M$ is chosen as discussed in Section~\ref{sec-choose-M}. This choice ensures that the 
%size 
area/volume 
of each pixel/voxel is less than $\varepsilon$ times the area/volume of the smallest grain, and thus depends on whether the APD is single-phase or multi-phase. 

%In this paper we use the term \emph{single phase} to refer to APDs with grains of equal volume, which represent idealised monodisperse microstructures, where are all the grains have essentially the same size. We use the term \emph{multi phase} to refer to polydisperse microstructures. 
%\csr{I've sometimes wondered about describing the monodisperse case as single phase. In materials science the phase refers to the composition and physiochemical properties, but here we're just referring to the fact that the grains are all the same size. Not sure whether to change it at this stage, but just thought I'd bring it up.}
%
In the single-phase examples the areas/volumes $V = (v_i)_{i=1}^N$ of grains are equal, $v_i = |\Omega|/N=1/N$ for all $i$. 
%As an example, in %this 
In this
case, when $D=3$ and $N=10,000$, setting the inverse pixel length parameter to $M = 200$ (which results in $M^D=8,000,000$ voxels) lets us achieve the 1\% tolerance reliably.
%\cdb{Where do we state the values of $M$ that we use in 2D and 3D for each value of $N$ for every runtime test? I can't find this anywhere.} 

%\cmb{Added a sentence above. Is that enough? I thought we agreed that the precise discussion about $M$ can be deferred and that just indicating that it has be "high enough for the problem at hand" is enough here? Note that the choice of $M$ is fluid for multi-phase random runs, since areas of smallest grains differ, but this makes perfect sense and is the smarter approach, so I do not think we should aim for explicitly stating the values of $M$, as it does not really add much? The general usual philosophy, as in, say, discretising pdes with finite elements, is that we want $M$ as small as possible whilst still being able to hit the tolerance. But one does not know a-priori what is the smallest choice of $M$ that would work for given data, so a heuristic guess is needed and the more conservative the guess is, the more likely it is that it will always work. Ours is pretty conservative and there has not been a single instance so far that it failed. At the same time, it does take data into account, so it varies.}

In the multi-phase examples the areas/volumes are drawn from a lognormal distribution with shape parameter $\sigma =1.0$ and location parameter $\mu = 0.5$ and subsequently normalised so that the total sum of the areas/volumes is $|\Omega|=1$. It follows from \eqref{choice-of-M} that for multi-phase problems, the value of $M$ has to be increased to reflect the size of the smallest grain. 
%Note that in the multi phase problem  are much more computationally challenging as they require $M$ to be  %Note that in this case, the lower bound on the choice of $M$ from \eqref{eqn-lower-bdd-M} becomes
%\[
%\frac{1}{M^D} < \varepsilon \min_i v_i,
%\]
%ensuring that each pixel/voxel is of size lower than $\varepsilon$ times the size of the smallest target mass. This makes the multi-phase examples much more computationally challenging. \cmb{rewrite this paragraph: simplify, mention that $M$ has to be sufficiently large  but not larger (to not increase the runtime), refer to Methods section for proper explanation} 

%\cmb{rewrite this paragraph to just inform about double/single precision choices for the tests} 

Single precision arithmetic is employed for runtime test (a) (computing fixed APDs via Algorithm~\ref{alg:pixel_APD}) and double precision arithmetic is used for runtime test (b) (finding optimal APDs via Algorithm~\ref{alg:optimal_APD}). The switch to double precision is
%{\db \sout{dictated by frequently observed precision loss.}}
necessary to avoid precision loss and to achieve the desired error tolerance $\varepsilon$.
%\cmb{Isn't this somewhat vague? I wouldn't say precision loss is a bit of a different beast altogether? We introduced the "desired error tolerance" and mentioned how $M$ has to be adjusted to have a chance of achieving it, but at no point prior to this sentence do we mention the effect of floating point arithmetic on it? }
%\cdb{How about this new version, which mentions both precision loss and accuracy? I think both are important? If $\varepsilon$ is very small, then single precision would not allow us to hit the desired accuracy $\varepsilon$, e.g., if the gap after the floating point number $v_i$ before the next floating point number is bigger than $\varepsilon$.}

%Based on extensive testing, we report that single precision arithmetic is reliable for computing the APDs via Algorithm~\ref{alg:pixel_APD}, but, except for small problems, the single-precision arithmetic leads to precision loss when finding an optimal APD via Algorithm~\ref{alg:optimal_APD}, in the sense that the solver reports reaching desired tolerance of 1\% area/volume error, but a manual check afterwards reveals errors exceed the tolerance. Such an issue does not arise in double precision arithmetic. 

To avoid random effects, in all runtime tests we report the mean runtime and the full range of observed runtimes over ten random runs.

\begin{figure}
    \centering
    \begin{subfigure}[c]{0.48\textwidth}
        \centering
        \subcaption{2D}
        \includegraphics[trim=0 0 0 30,clip,width=\textwidth]{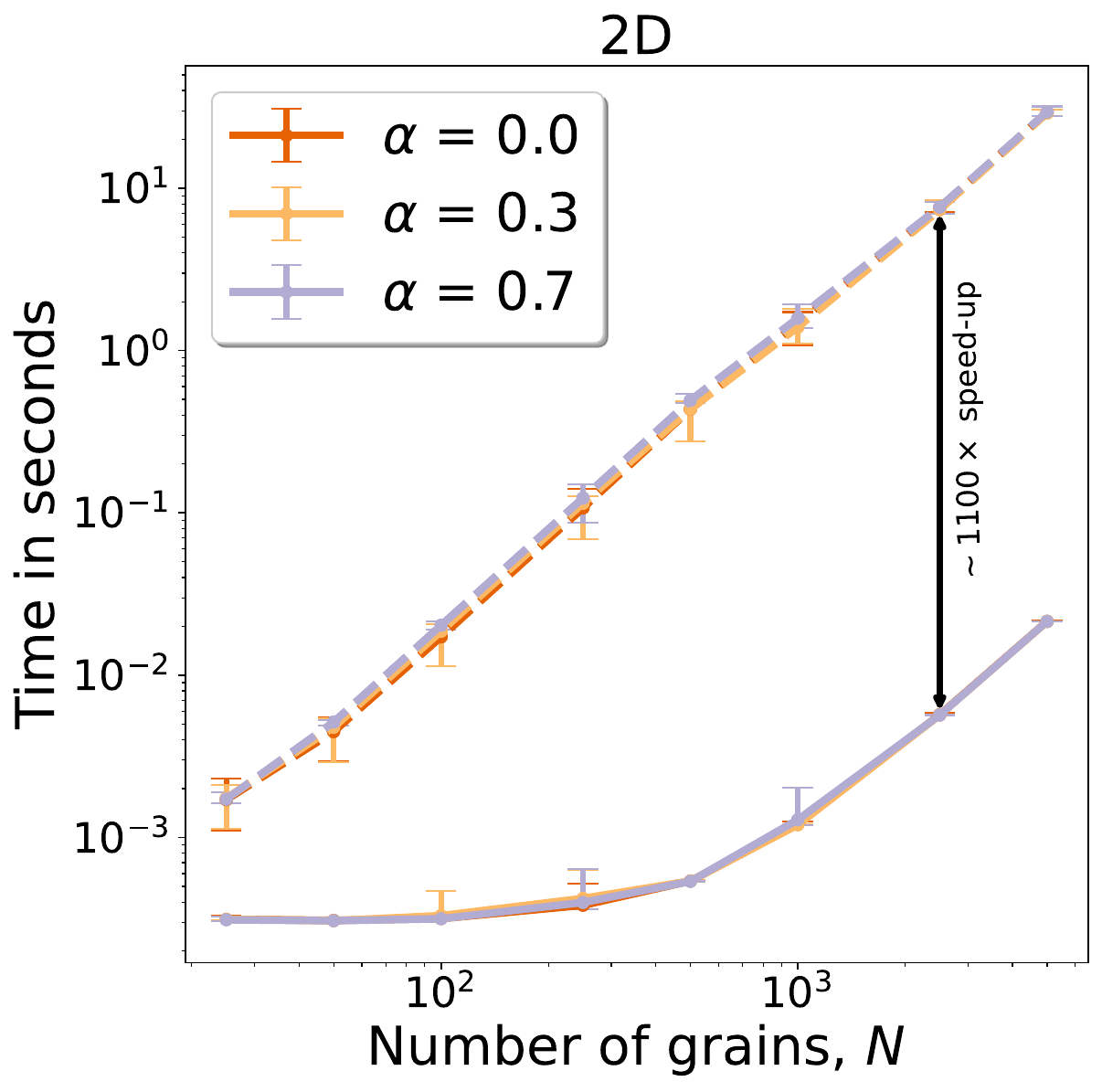}
        %\caption{EBSD scan}
        %\label{fig:y equals x}
    \end{subfigure}
    %\hfill
    \begin{subfigure}[c]{0.48\textwidth}
        \centering
        \subcaption{3D}
        \includegraphics[trim=0 0 0 30,clip,width=\textwidth]{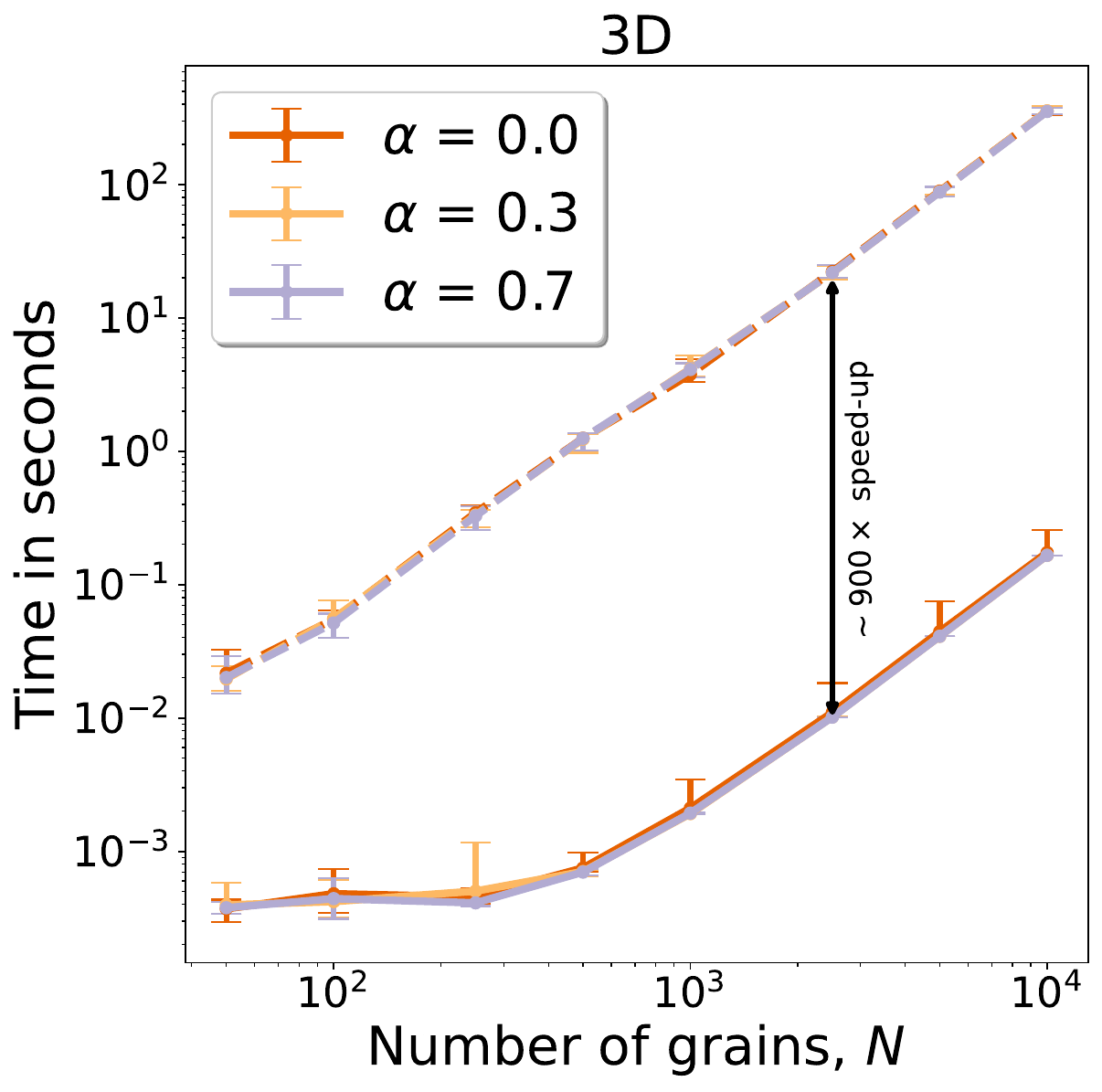}
        %\caption{Fitted centroids and ellipses.}
        %\label{fig:three sin x}
    \end{subfigure}
    \caption{Runtime test (a). We use %using 
Algorithm~\ref{alg:pixel_APD} to compute 
%an 
APDs in 2D (A) and 3D (B), with weights $W$ set to zero and seeds $X$ and anisotropy matrices $\Lambda$ sampled randomly as described at the start of Section~\ref{sec:runtime}. The runtimes are averaged over 10 random runs.
The computation was performed on a single A100 GPU with single precision (solid lines) and compared with a reference run on a CPU (dashed lines). The GPU computation was about 1000-times faster than the CPU computation.
By checking against the same computation in double precision, we have verified that even for the largest problems employing single precision arithmetic to compute an APD does not lead to precision loss and the miss-assignment of pixels is minuscule. Note that the runtimes are essentially independent of the anisotropy parameter $\alpha$.}
       \label{fig:runtime-apd-gen}
\end{figure}

\begin{figure}
    \centering
    %\vskip\baselineskip
    \begin{subfigure}[c]{0.34\textwidth}
        \centering
        \includegraphics[trim=0 0 0 0,clip,width=\textwidth]{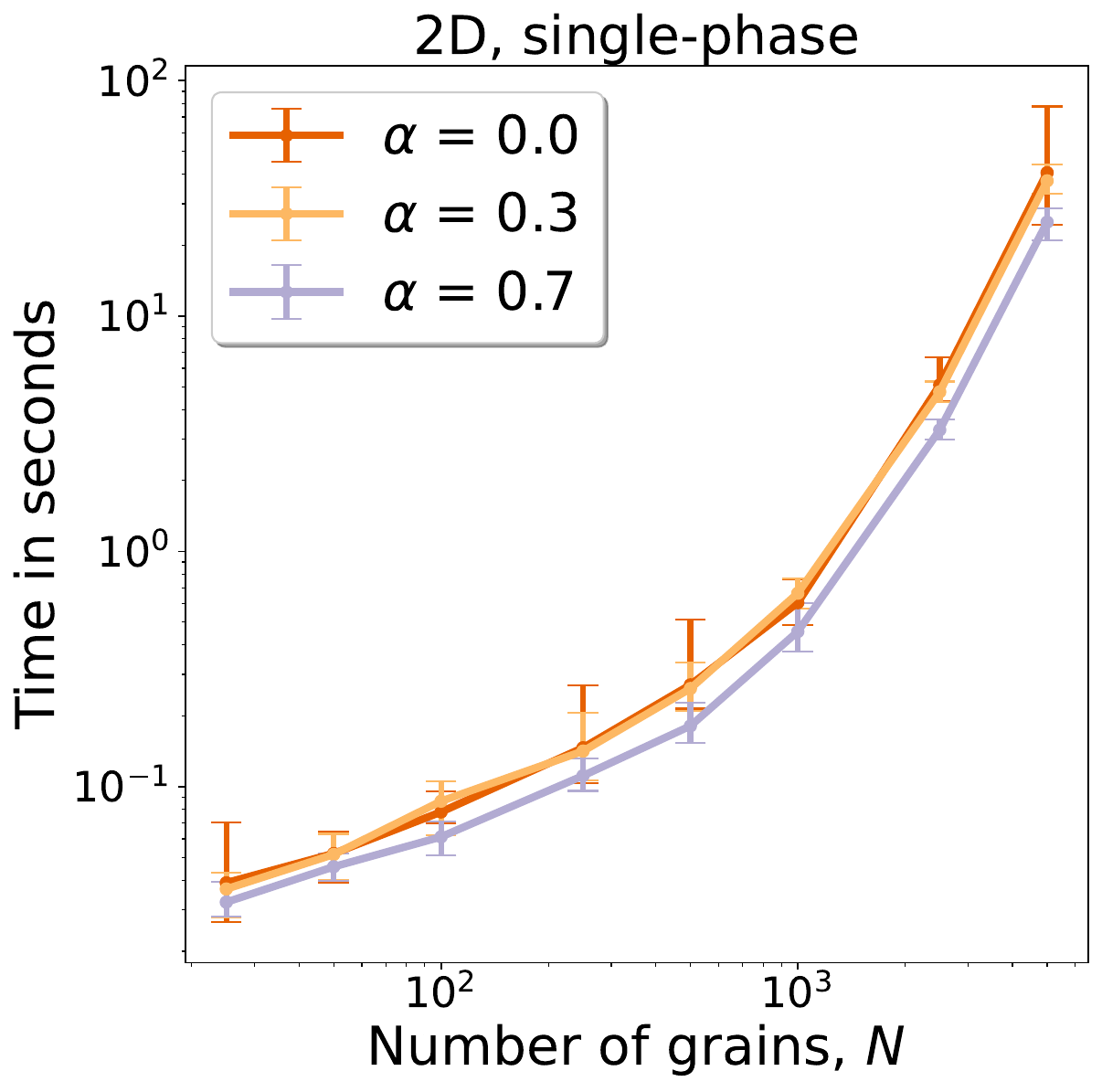}
    \end{subfigure} %\hfill
    \begin{subfigure}[c]{0.34\textwidth}
        \centering
        \includegraphics[trim=40 40 40 80,clip,width=\textwidth]{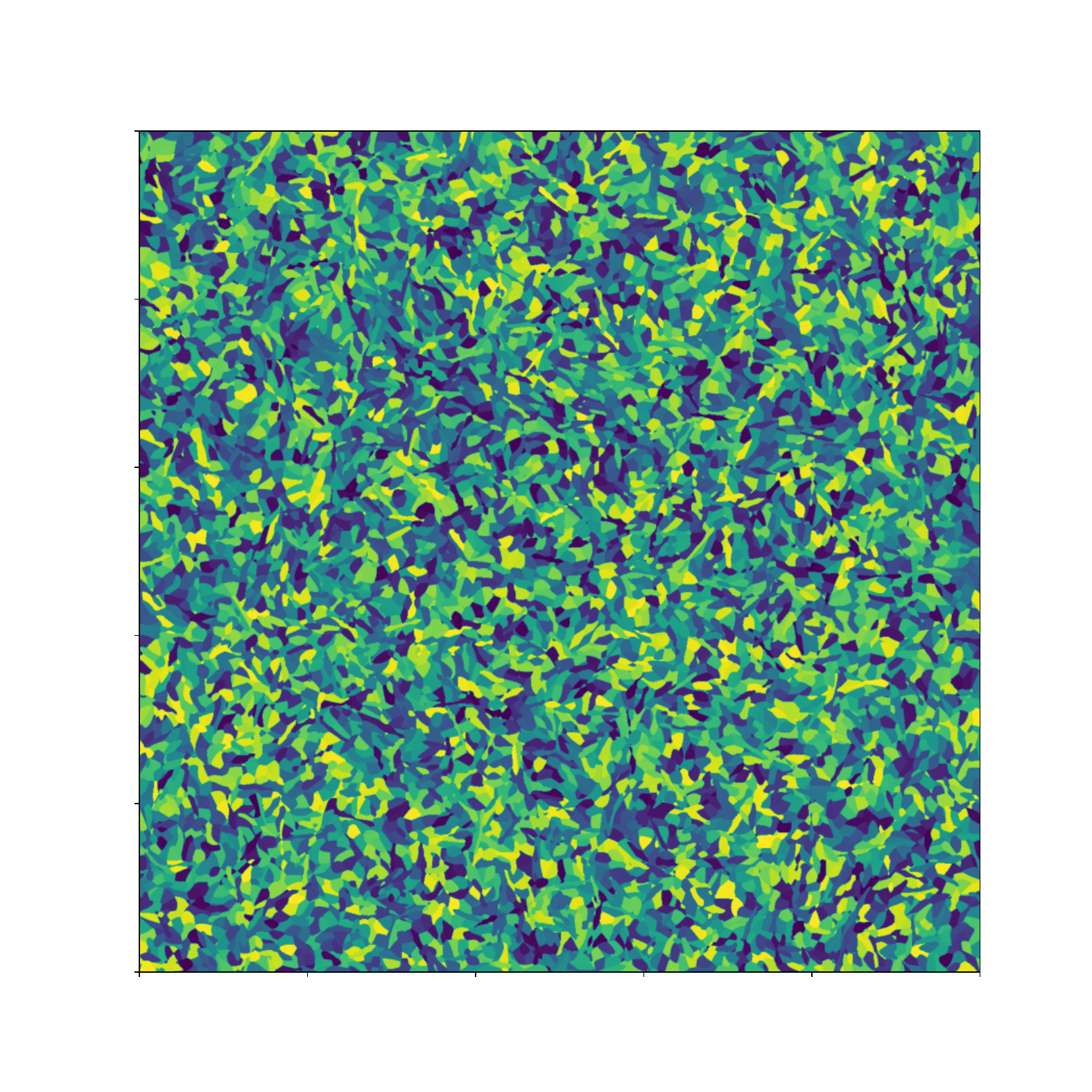}
    \end{subfigure}
    \begin{subfigure}[c]{0.34\textwidth}
        \centering
        \includegraphics[trim=0 0 0 0,clip,width=\textwidth]{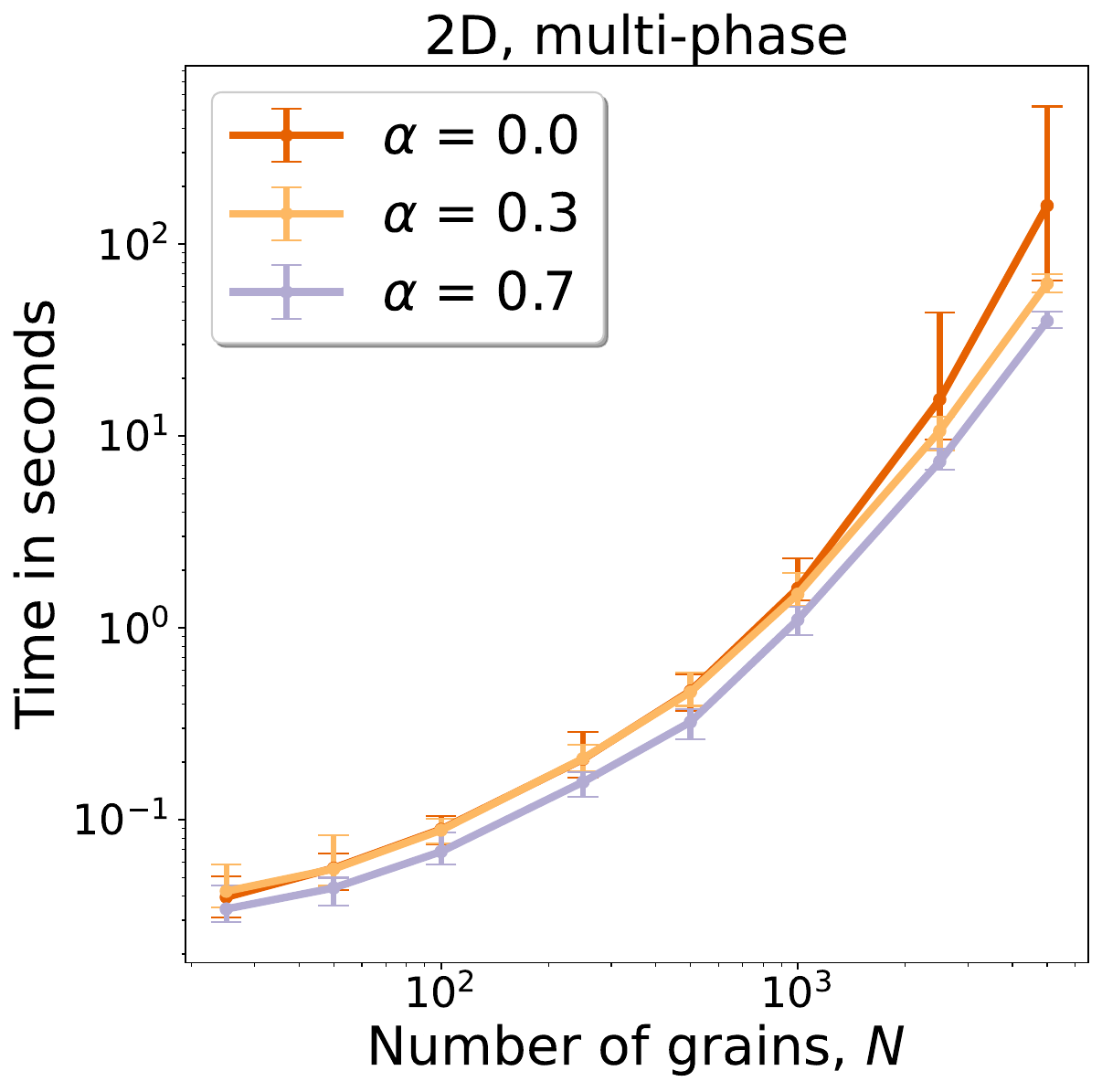}
    \end{subfigure}
    \begin{subfigure}[c]{0.34\textwidth}
        \centering
        \includegraphics[trim=40 40 40 80,clip,width=\textwidth]{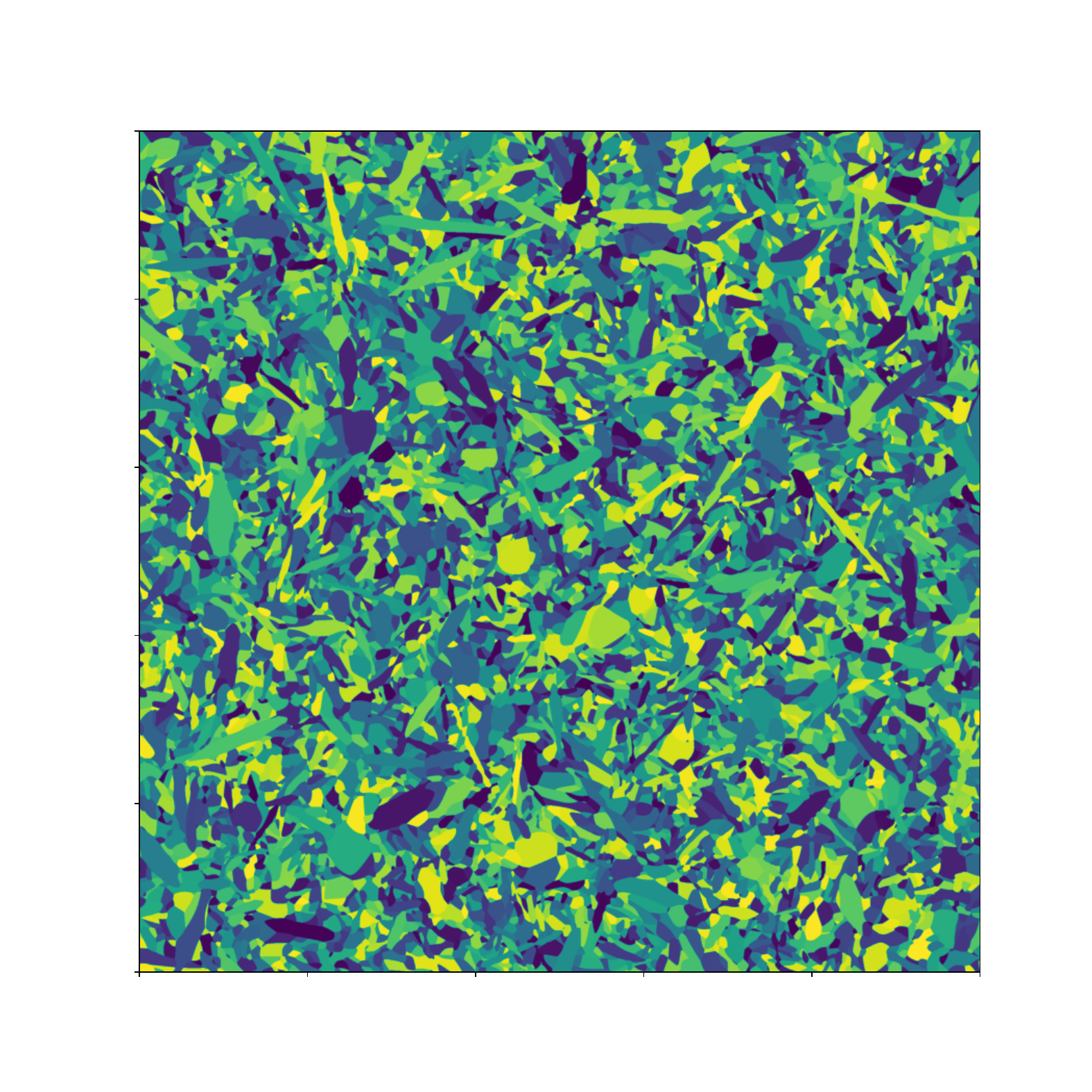}
    \end{subfigure}
    \begin{subfigure}[c]{0.34\textwidth}
        \centering
        \includegraphics[trim=0 0 0 0,clip,width=\textwidth]{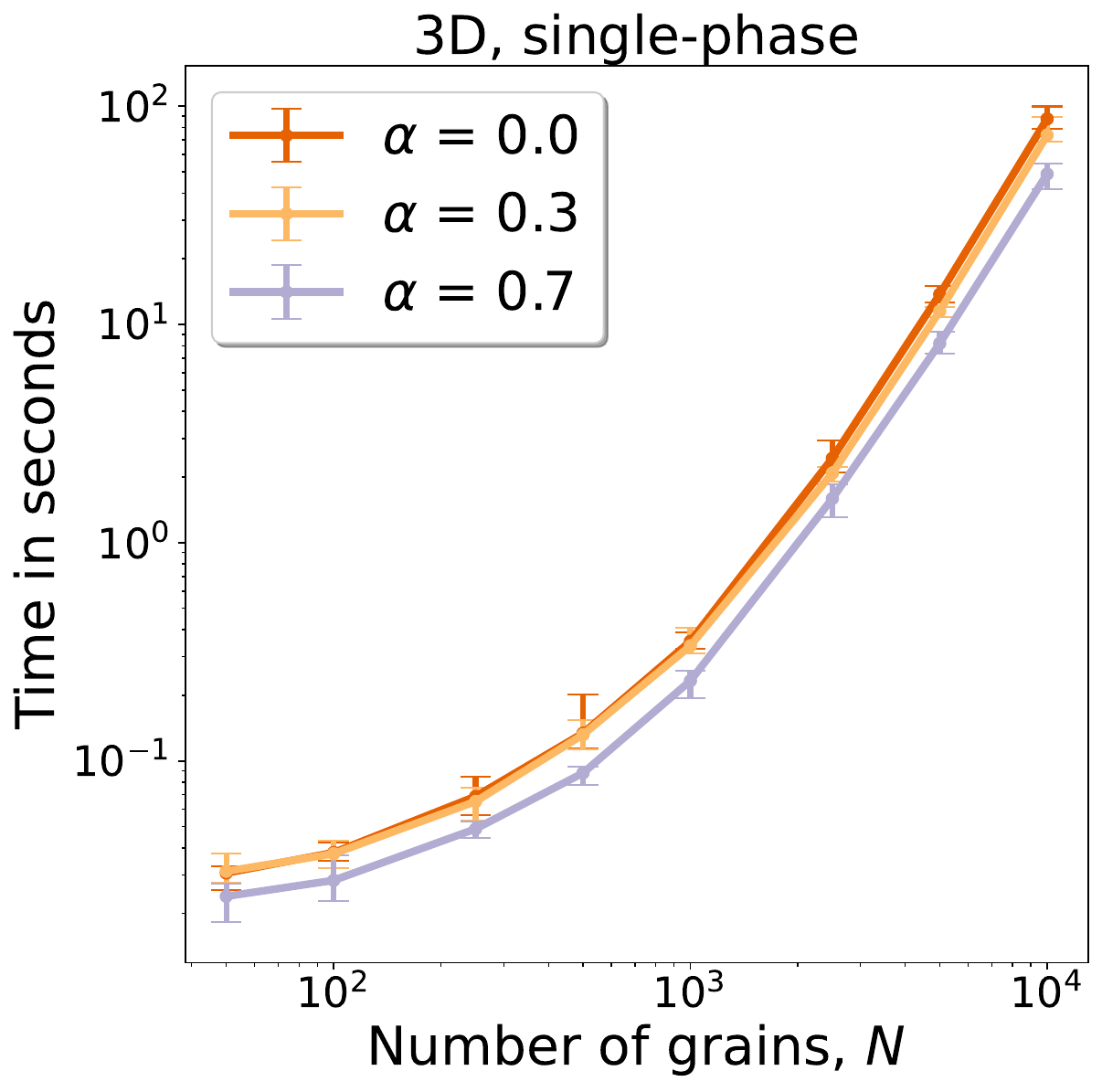}
    \end{subfigure}
    \begin{subfigure}[c]{0.34\textwidth}
        \centering
        \includegraphics[trim=500 0 500 200,clip,width=\textwidth]{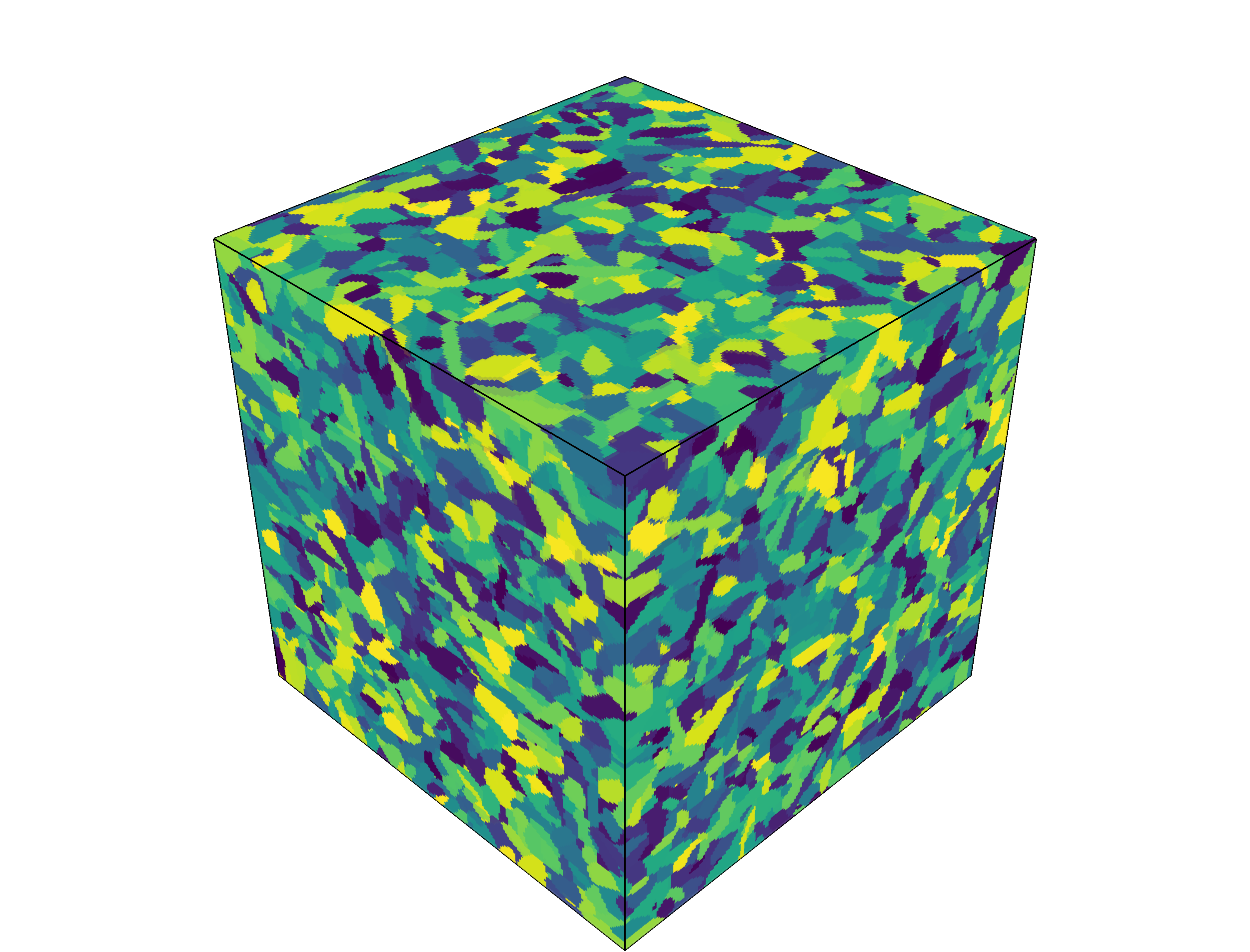}
    \end{subfigure}
    \begin{subfigure}[c]{0.34\textwidth}
        \centering
        \includegraphics[trim=0 0 0 0,clip,width=\textwidth]{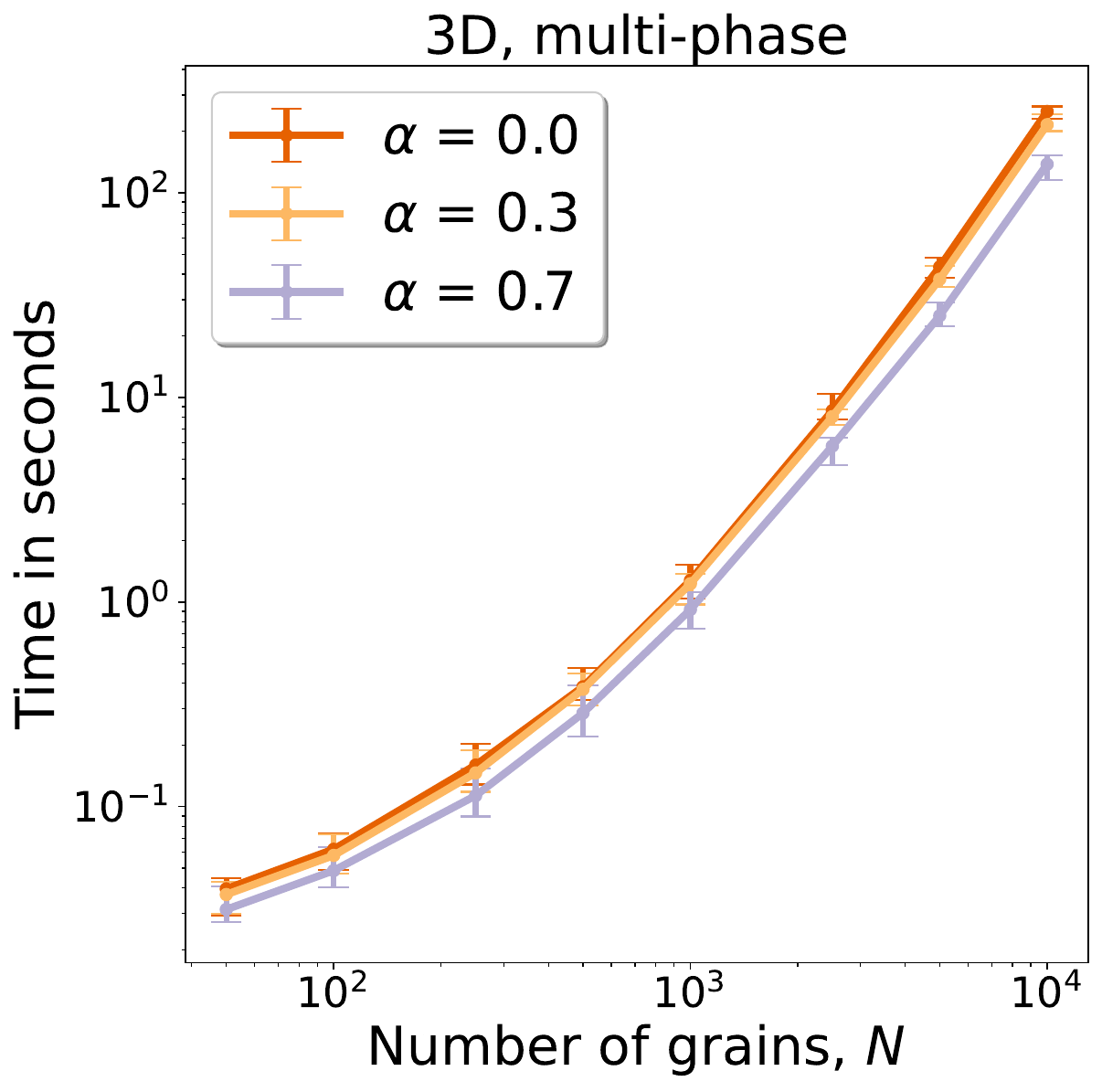}
    \end{subfigure}
    \begin{subfigure}[c]{0.34\textwidth}
        \centering
        \includegraphics[trim=500 0 500 200,clip,width=\textwidth]{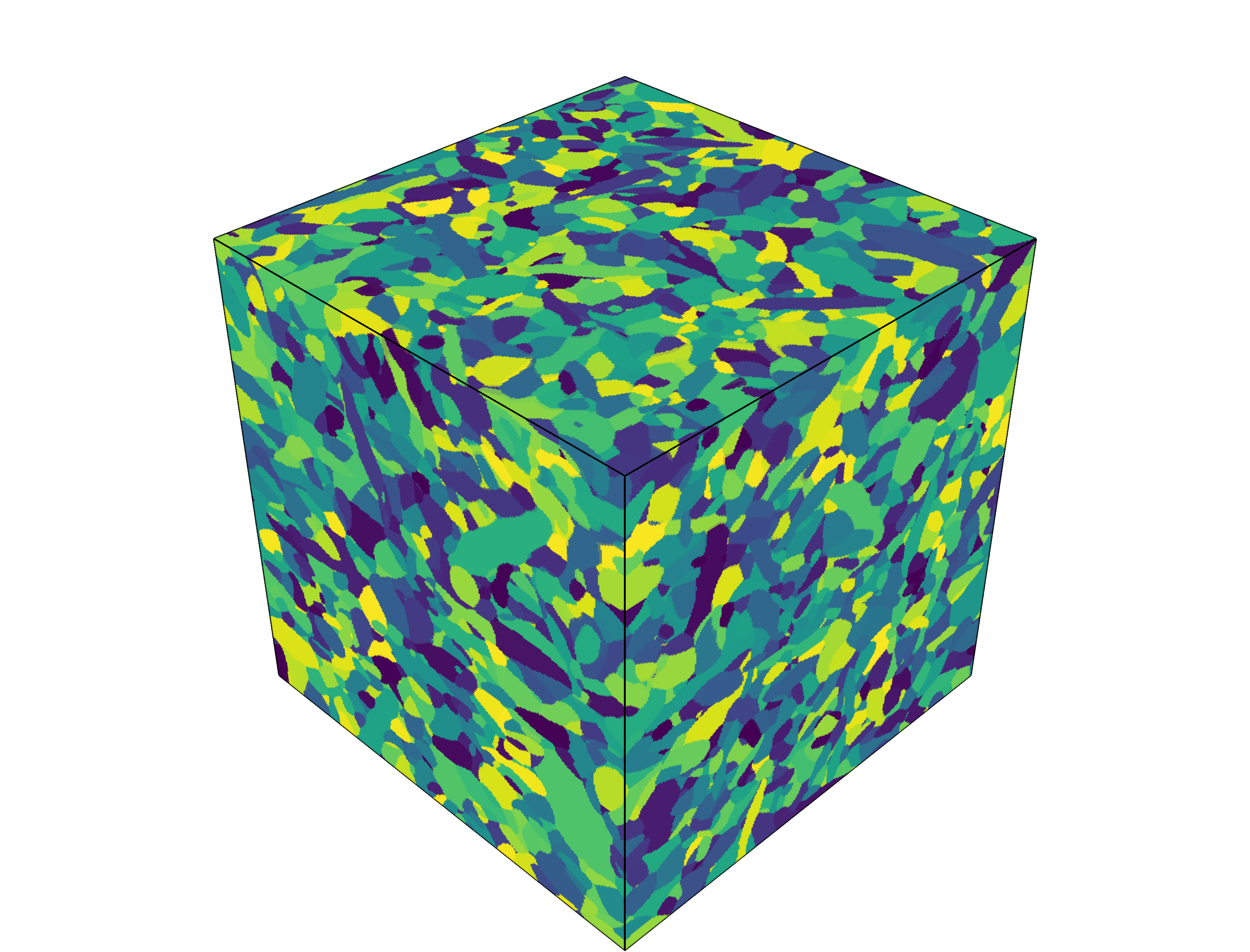}
    \end{subfigure}
    \caption{Runtime test (b). We use Algorithm~\ref{alg:optimal_APD} to find optimal APDs with cells of prescribed volumes on a single A100 GPU with double precision, in 2D (top two rows) and 3D (bottom two rows), single phase (1st and 3rd rows) and multi phase (2nd and 4th rows), as described in Section \ref{sec:runtime}. On the right we plot optimal APDs when $N=5000$ (in 2D) and $N=10,000$ (in 3D) with anisotropy parameter $\alpha = 0.7$, with colours, in the absence of crystallographic data, assigned randomly. The runtimes are averaged over 10 random runs. The areas/volumes of the grains are accurate to $1\%$.}
       \label{fig:runtime-optimal-apds}
\end{figure}

The results are presented in Figure~\ref{fig:runtime-apd-gen} and Figure~\ref{fig:runtime-optimal-apds}. 
Notably, with our implementation we are able to maintain a near 100\% GPU usage throughout. Hence, if these tests were run on different GPUs, the relative timing difference would closely follow the relative differences in performance reported in Table~\ref{table:FLOPS}. It is also for this reason that our implementation is so fast. To make this point clear, for runtime test (a) we also report runtimes
%for test (a) 
in a CPU-only setup, performed on the AMD EPYC 7713 CPU, again provided by NERSC. The algorithm runs about 1000 times faster on the GPU than on the CPU.

%\cdb{We should make some comments on what we conclude from the results, I don't think it's enough to simply report them without comment. Perhaps something along the following lines?}
From Figure \ref{fig:runtime-apd-gen} we see that we can generate APDs with $5000$ grains in 2D in the order of $10^{-2}$ seconds, and $10,000$ grains in 3D in the order of $10^{-1}$ seconds. From Figure \ref{fig:runtime-optimal-apds} we observe that we can generate multi-phase, anisotropic ($\alpha >0$), optimal APDs (with grains of prescribed volumes) with $5000$ grains in 2D in about $1$ minute or less, and $10,000$ grains in 3D in the order of $10^{2}$ seconds.
The runtime for multi-phase APDs is longer than that for single-phase APDs, as expected. Surprisingly, the runtime \emph{decreases} as the anisotropy parameter $\alpha$ \emph{increases}. In other words, it is slower to compute isotropic power diagrams than anisotropic diagrams using our method. This is not a problem, however, since for isotropic power diagrams (where the cells are much simpler, namely convex polytopes with flat boundaries) there are much faster algorithms and implementations, such as \cite{KMT19,BKRS20,KSSB20}. For example, a 3D multi-phase, optimal \emph{isotropic} power diagram with $10,000$ grains of prescribed volumes can be computed in less than 20 seconds on a standard CPU laptop \cite{BPR23,LPM}, and even faster implementations exist, such as \cite{pysdot} and \cite{Geogram}. However, these methods do not apply to \emph{anisotropic} power diagrams, for which there is currently no faster alternative to our library \textsc{PyAPD} \cite{PyAPD}, as far as we are aware. 
%\cmb{this paragraph is very nice!}
%\cdb{Thanks!}

%\begin{example}[Runtime test: APD computation]\label{ex:runtime_APD}
%    We consider generating an approximate APD for a given triplet  $(X,\La,W)$. 
    % single phase optimal APDs with $N$ grains inside $\Omega = [0,1]^2$, with 
    %\begin{equation}\label{eqn-N_choice_2D}
    %N = 25, 50, 100, 250, 500, 1000, 2500, 5000.
    %\end{equation}
    %using uniform discretisation with inverse pixel length parameter $M=1200$, resulting in $M^2 = \num[group-separator={,}]{1440000}$ pixels. This is roughly in line with the lower bound on the choice $M$ presented in Table~\ref{table:choose_M}. The results are presented in Figure~\ref{fig:runtime1a} and Figure~\ref{fig:runtime1b}.
%\end{example}

%\begin{example}[Runtime test: finding optimal weights]\label{ex:runtime2}  
%In the single phase examples the areas/volumes $V = \{v_i\}_{i=1}^N$ of grains are equal and thus set to $v_i = 1/N$. In the multi phase examples the areas/volumes are drawn from a lognormal distribution with shape parameter $\sigma =1.0$ and location parameter $\mu = 0.5$ and subsequently normalised so that the total sum is 1.
%\end{example}

\subsection{Fitting APDs to EBSD measurements}\label{sec:ex_ebsd} 
In the following example, we work with an experimentally measured microstructure obtained using the EBSD technique specifically for this study. The initial microstructure and crystallographic texture of the material were measured across the thickness (ND - normal direction) perpendicular to the rolling direction (RD). We performed the EBSD measurements on an area located at the mid-thickness of the rolling plane (ND-RD plane). The EBSD scan area is \SI{901.5}{\micro\metre} $\times$ \SI{999.52}{\micro\metre}, and it  was measured with a step size of 1.0 and 0.85 \SI{}{\micro\metre} in the rolling and normal directions, respectively. This results in 1,039,754 pixels (901 $\times$ 1154). Standard metallographic techniques were used to prepare the specimen for characterisation. Analysis of the EBSD data was performed using the TSL OIM software. The material used in the present study is a low carbon steel.

Following a standard postprocessing procedure done in the \textsc{MTEX} toolbox \cite{MTEX}, we obtain a grain file of ${N=4587}$ grains containing information about the areas of the grains 
%$V = \{v_i\}_{i=1}^N$; 
$V = (v_i)_{i=1}^N$;
the locations of the centroids of the grains 
%$X = \{x_i\}_{i=1}^N$; 
%$X = (x_i)_{i=1}^N$;
$(c_i)_{i=1}^N$;
the major and minor axes and orientations of the ellipses best describing the anisotropy of the grains, thus giving rise to a set of anisotropy matrices %$\La = \{\mathbf A_i\}_{i=1}^N$. 
$\La = (\mathbf A_i)_{i=1}^N$.
We note here that the ratio of areas $\tfrac{\max_i v_i}{\min_i v_i} > 188$, which makes reaching our target accuracy for the smallest grains particularly challenging. The original EBSD file, the script for postprocessing in \textsc{MTEX} and the resulting grain file are available through the library \textsc{PyAPD} \cite{PyAPD}. 

%Based on this, 
We fit an optimal APD to the grain file data as follows.
%\cdb{More details needed, something along the following lines:}
We take the seeds $x_i$ of the APD to be the centroids returned by \textsc{MTEX}, $x_i = c_i$ for all $i$. Similarly, we take the anisotropy matrices $\mathbf{A}_i$ of the APD
to be those returned by \textsc{MTEX}. The weights $w_i$ are found using Algorithm \ref{alg:optimal_APD}, to ensure that the APD cells $L_i$ have areas $v_i$ (given by \textsc{MTEX}) up to the  relative error tolerance  $\varepsilon = 0.01$. 
%\cmb{I don't understand, this is described in the previous paragraph? Apart from how we find the weights, but for this we just need to mention Algorithm~\ref{alg:optimal_APD}, which the old wording contained}
%\cdb{It's not described in the previous paragraph. The previous paragraph explains what numbers MTEX extracts from the data. Whereas this paragraph explains how we choose the generators of the APD. We're mixing up two different steps. I've introduced different notation for the centroids ($c_i$) so that it doesn't look like repetition.}
%The optimal APD is obtained using Algorithm~\ref{alg:optimal_APD}, with the  relative error tolerance set to $\varepsilon = 0.01$ and 
The inverse pixel length parameter $M$ is chosen according to \eqref{choice-of-M}.

\begin{figure}
    \centering
    \begin{subfigure}[c]{0.48\textwidth}
        \centering
        \subcaption{}
        {{\includegraphics[trim = 40 0 40 0,clip,width=\textwidth]{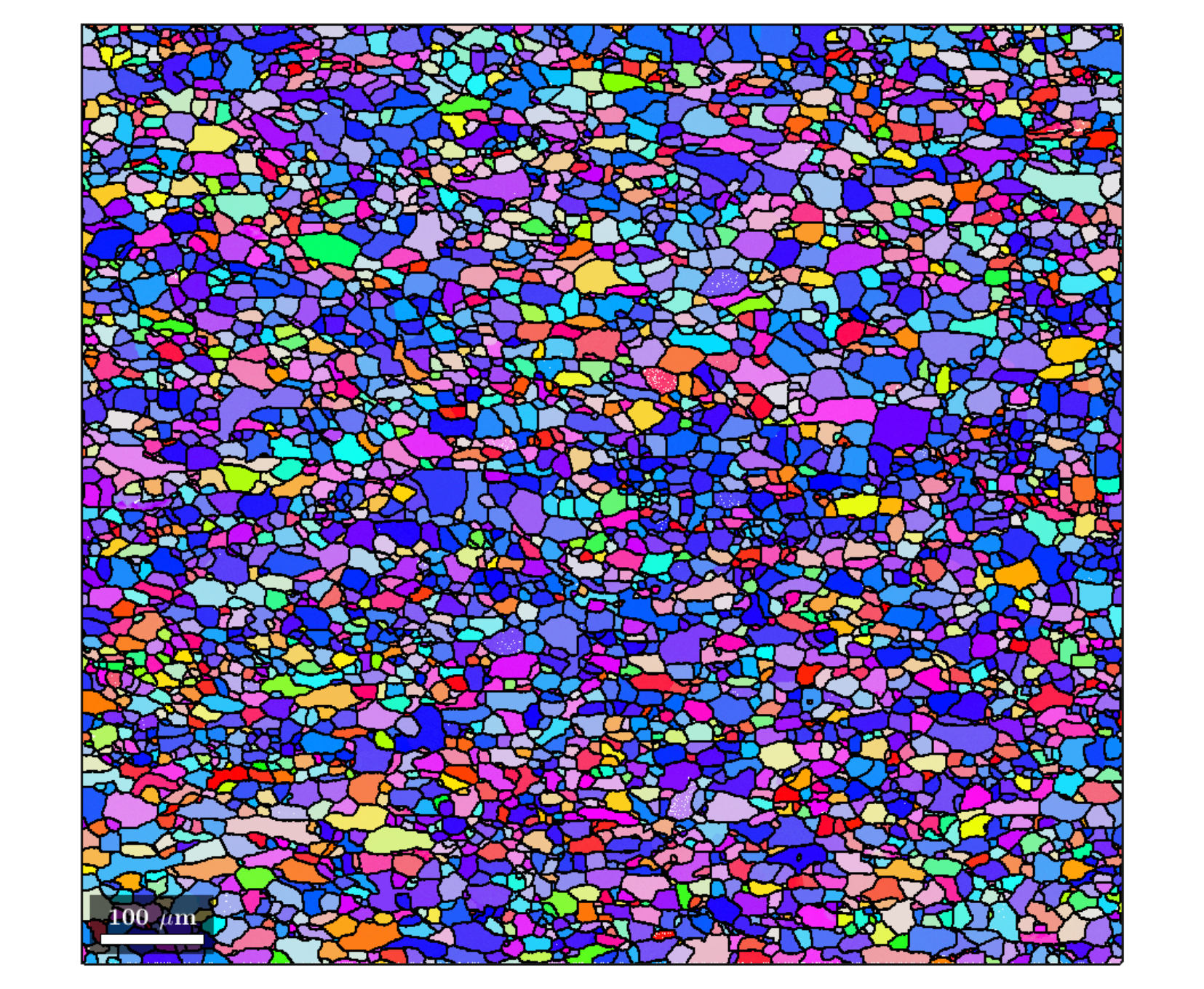}}\llap{\includegraphics[trim=0 0 110 100,clip,height=2cm]{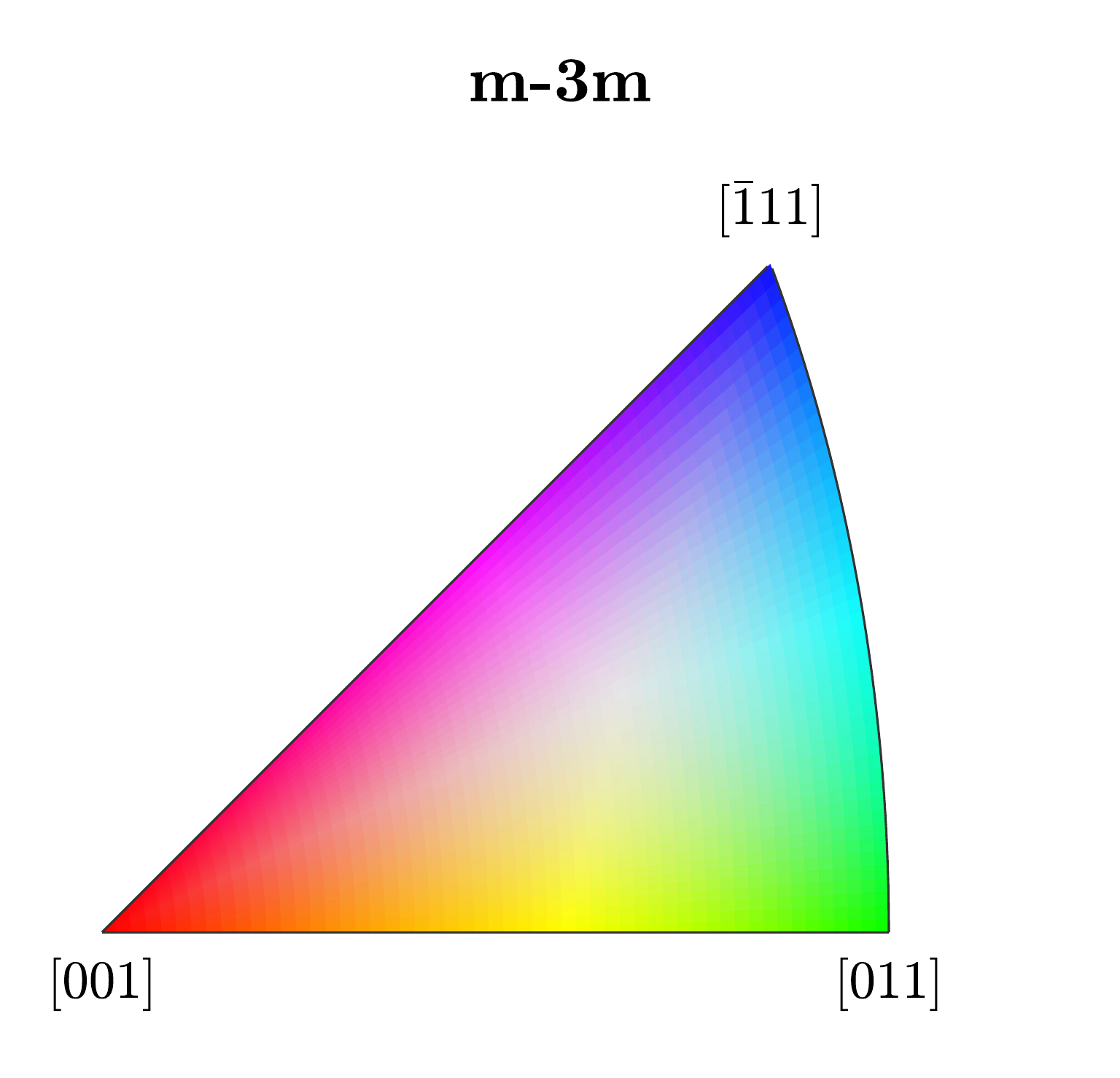}}}
        
        %\label{fig:y equals x}
    \end{subfigure}
    \hfill
    \begin{subfigure}[c]{0.48\textwidth}
        \centering
        \subcaption{}
        {{\includegraphics[trim = 40 0 40 0,clip,width=\textwidth]{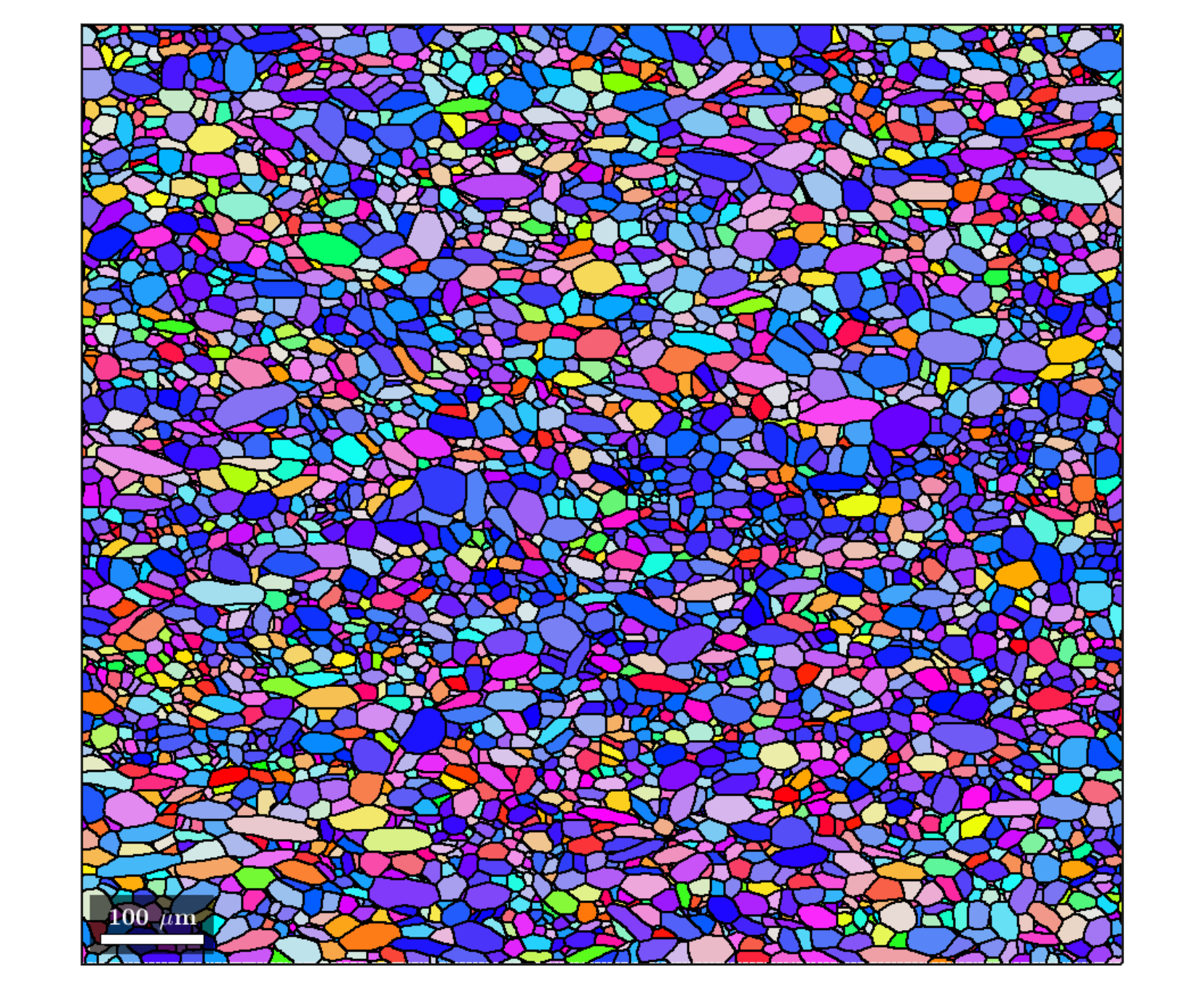}}}
        %\label{fig:y equals x}
    \end{subfigure}
    \begin{subfigure}[T]{0.43\textwidth}
         \centering
         \subcaption{}
         \scalebox{1.1}[1]{{\includegraphics[trim=0 0 0 0,clip,width=\textwidth]{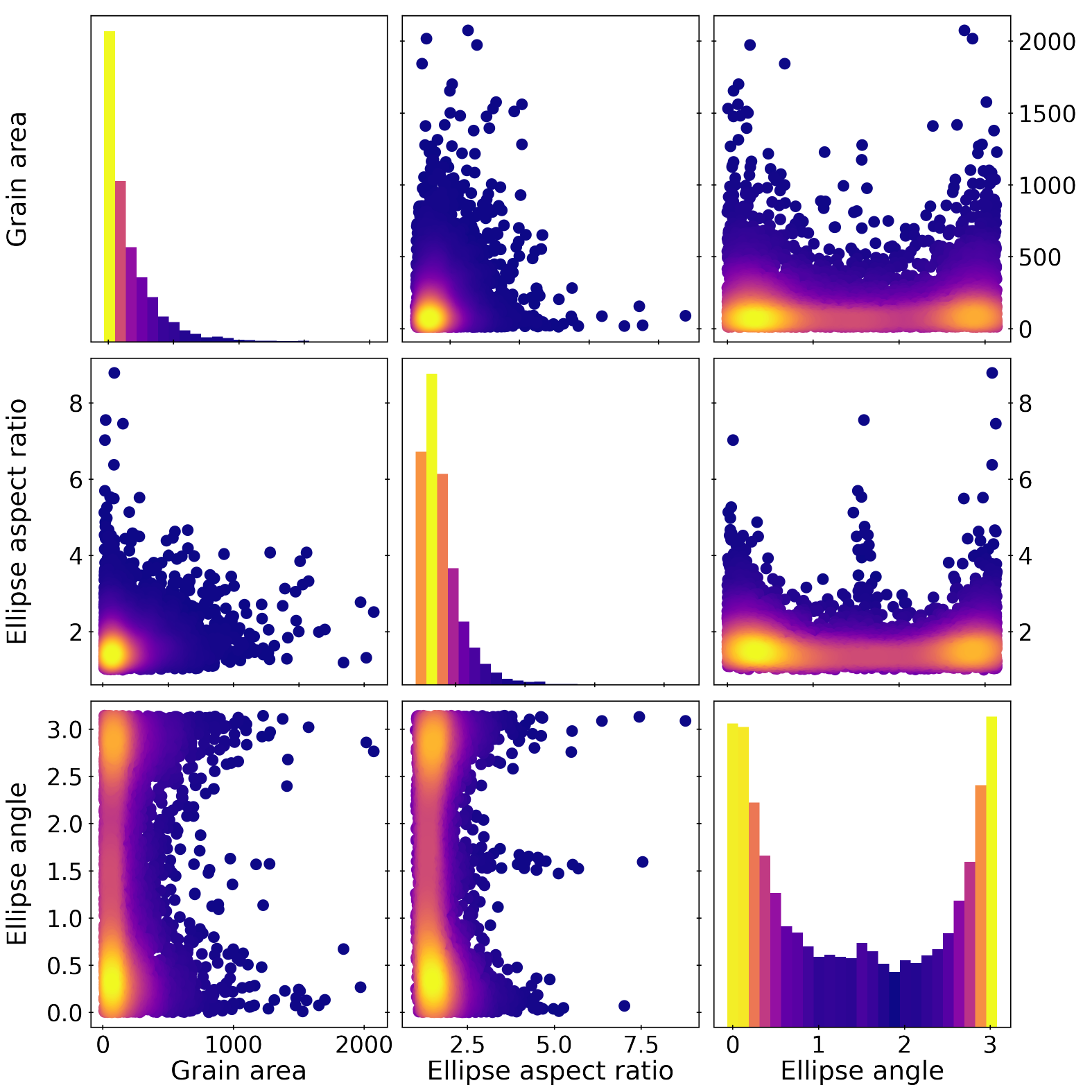}}}
     \end{subfigure}
     \hfill
    \begin{subfigure}[T]{0.485\textwidth}
        \centering
        \subcaption{}
        \scalebox{1}[1]{{\includegraphics[angle=0,trim=120 110 108 110,clip,width=\textwidth]{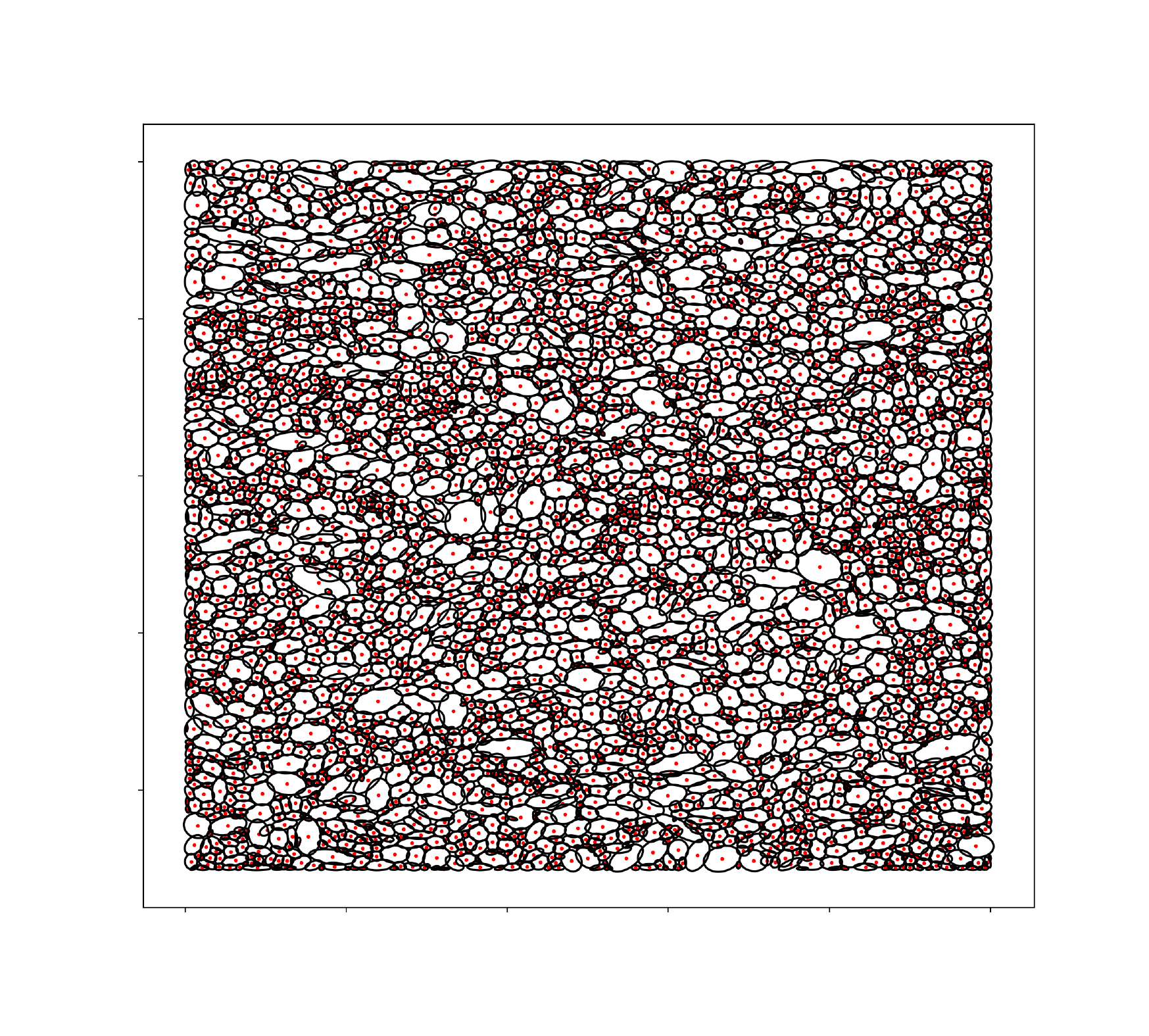}}}        
    \end{subfigure}
    \caption{%\cdb{Reworded to make it more accessible to mathematicians, while hopefully retaining the correct engineering terminology. Perhaps Karo could check this?} %{\db The IPF color maps parallel to the loading (vertical) direction for:} \cdb{I don't know what the previous sentence means. It looks like the sentence has been cut after ``for"?} \cmb{This was written by Karo, perhaps he can comment? I don't quite remember but at the time of reading it I felt like previous wording made sense, but it looks it has been erased now?} 
    %\cdb{I didn't erase anything after ``for". I think the ``for" refers to figures (A) and (B).} \cdb{I wonder if we should write it in simpler language for non-engineers, i.e., state what an IPF color map is? For mathematicians it would be enough to say that the grains have been coloured according to their crystallographic orientation.}
    {Fitting an optimal APD to EBSD data (see Section \ref{sec:ex_ebsd}).
    (A): The original EBSD scan. The grains are coloured according to their crystallographic orientation. To be precise, this is the 
     IPF color map parallel to the loading (vertical) direction for the original EBSD data.
    %and 
    (B): The corresponding IPF color map for the optimal APD. The relative error tolerance is $\varepsilon = 0.01 = 1\%$, and the APD took \SI{50.5}{\second} to generate.} (C): A scatter matrix plot of the statistical grain level data from the postprocessed EBSD data. \emph{Ellipse aspect ratio} refers to $a/b$ and \emph{ellipse angle} refers to $\theta$ in equation \eqref{eq: A}. (D): The postprocessed data illustrated in the form of centroids (red dots) and ellipses.}
       \label{fig:ebsd_fit1}
\end{figure}

The results are presented in Figure~\ref{fig:ebsd_fit1}.
%Using the optimal set of weights $W$ we obtain, we have checked against the pixelated EBSD data and found that our optimal APD
The optimal APD that we obtain is 89.53\% accurate, in the sense that this is the proportion of pixels that are assigned to the correct grain, while achieving a 1\% deviation in terms of the areas of the grains. For reference, the heuristic guess \cite{TR18} achieves 89.83\% pixel-level accuracy, 
 %which is broadly in line with literature \cite{AFGK23B}, 
 but the relative
 %volume 
 error of the areas is 380\%. This is also the reason why the heuristic guess is not always a great initial guess for Algorithm~\ref{alg:optimal_APD} - most of the
 computation
% {\db \sout{compute}}
% {\db computation} \cmb{these days people more often use "compute time", I was surprised at first but apparently with the advent of GPUs compute has become a noun :) }
% \cdb{We shouldn't use the word `compute' incorrectly just because other people do!}
 time is spent getting the areas of the small grains right. The proportion of the pixels that are correctly assigned could be increased by optimising the choice of $x_i$ and $\mathbf{A}_i$, rather than taking them directly from the data, as in for example \cite{AFGK23}, where accuracies of 93-96$\%$ are reported (for a different data set),  but this comes at a greater computational cost.%\cmb{What about the previous reference that was here, \cite{AFGK23B}?} \cdb{I added a reference to that earlier instead, on page 3, since I wanted to refer specifically to the numbers reported in \cite{AFGK23}.}

\subsection{Generating realistic synthetic microstructures}\label{sec:ebsd-synthetic}

We now turn our attention to generating 
%{\db \sout{samples of artificial representative volume elements}}
artificial microstructures
%\cdb{just to be a bit more succinct} 
that are statistically equivalent to the EBSD data from Section~\ref{sec:ex_ebsd} with respect to the joint distribution of the aspect ratios $a_i/b_i$, orientations $\theta_i$, and areas %{\db \sout{/volumes}}
$v_i$ of the grains.
%\cdb{We're working in 2D, so $v_i$ are areas, not volumes.}

To estimate the joint distribution of the grains in the grain file data, we use the multivariate variant of kernel density estimation 
%{\db \sout{, as}}
implemented in \textsc{OpenTurns} \cite{OpenTurns}. The fit and the scatter matrix plot in Figure~\ref{fig:ebsd_fit1} both reveal the statistical dependence between the grain properties. In particular, we observe that high anisotropy is mostly observed for grains with small area/volume. %{\db \sout{, which intuitively makes sense -- long and thin shapes have smaller areas than round shapes.}}
%\cdb{I'm not sure if this is true. Long and thin shapes can have larger areas than round shapes. Or perhaps you mean `long and thin shapes have smaller area than round shapes with the same diameter, by the  isodiametric inequality'?}

\begin{figure}
     \centering
     \begin{subfigure}[c]{0.47\textwidth}
         \centering
         \subcaption{}
         \scalebox{1}[1.15]{\includegraphics[trim=0 0 0 10,clip,width=\textwidth]{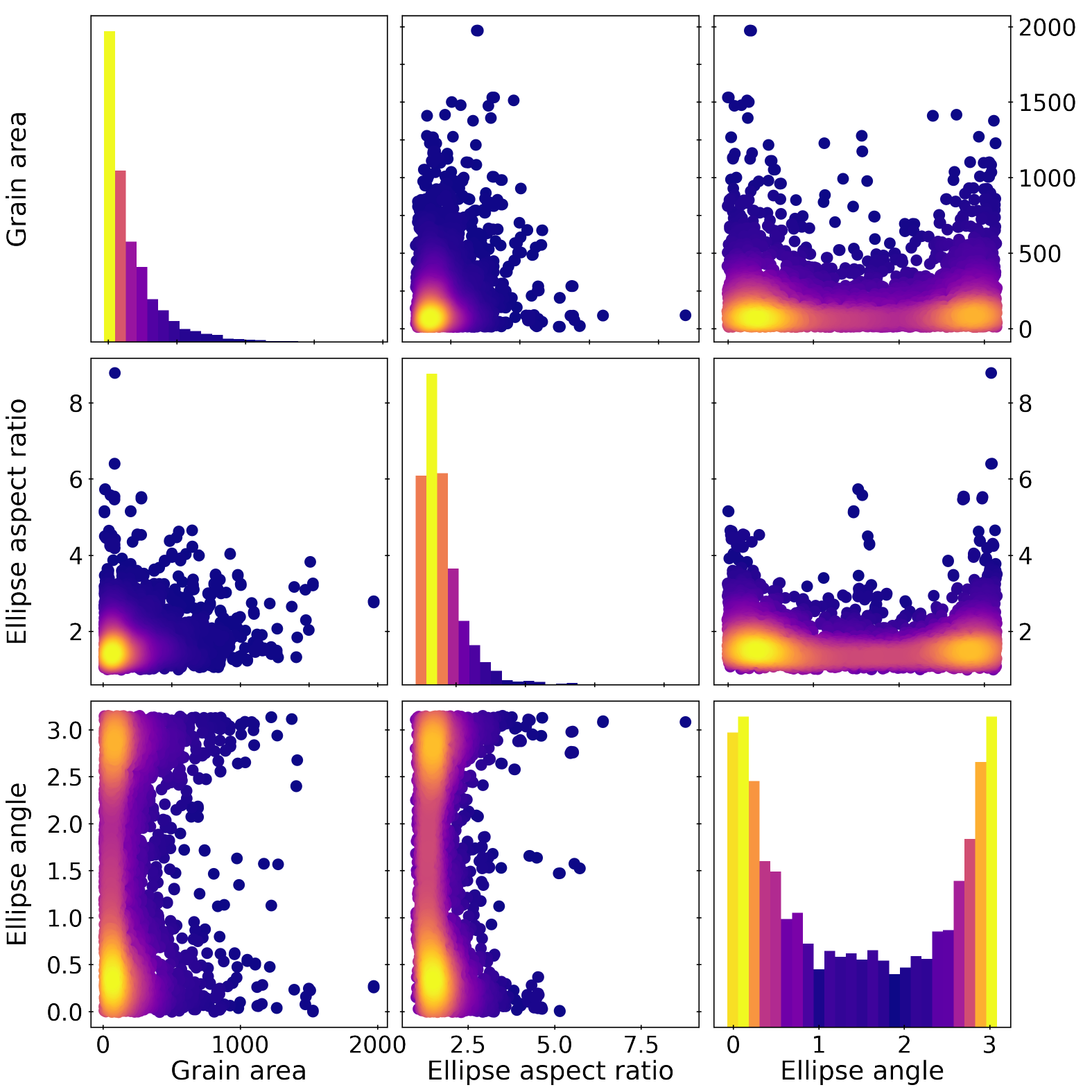}}
         %\caption{Optimal APD.}
         %\label{fig:five over x}
     \end{subfigure}
     \hfill
     \begin{subfigure}[c]{0.5\textwidth}
         \centering
         \subcaption{}
         \scalebox{1}[1]{{\includegraphics[angle =90 ,trim=100 105 90 120,clip,width=\textwidth]{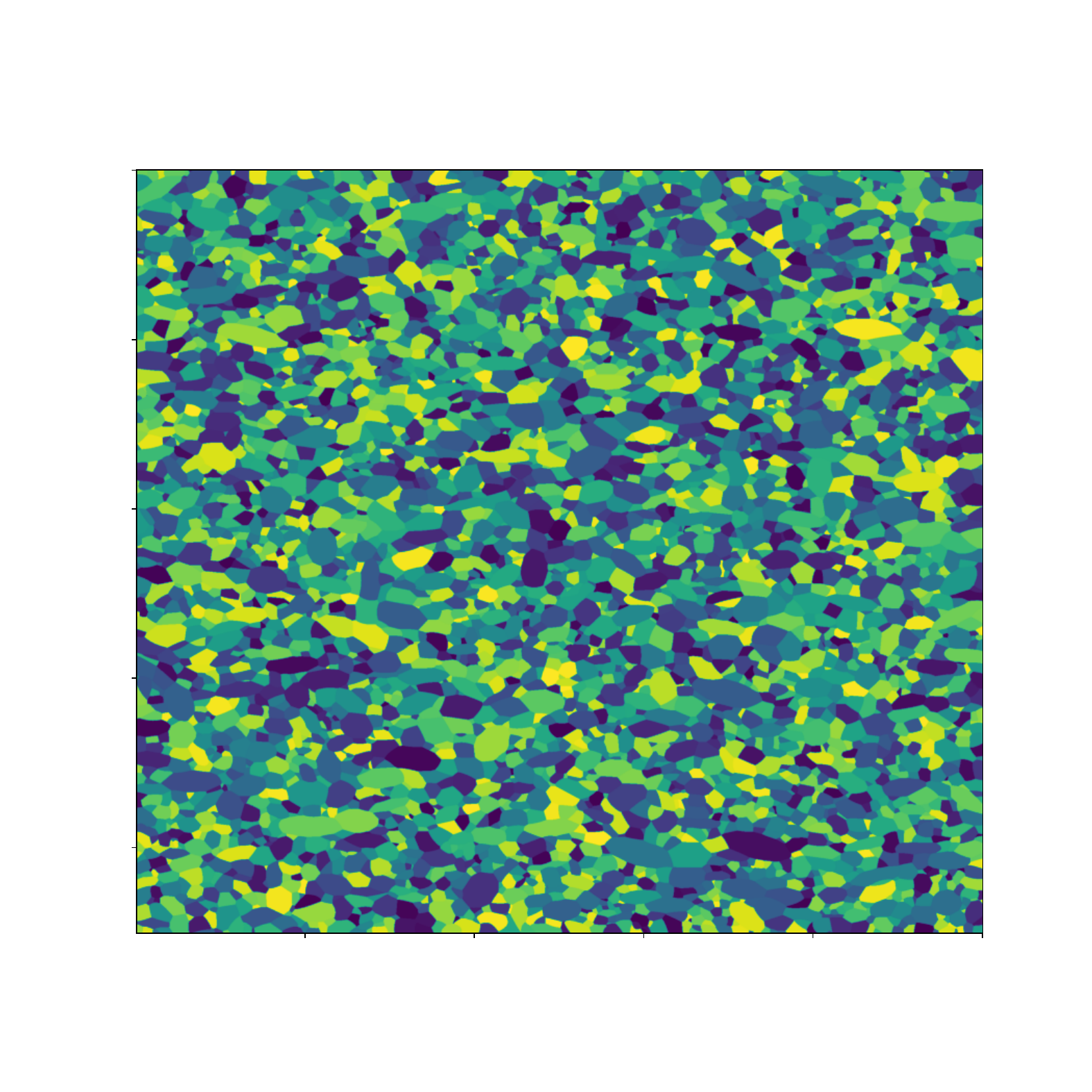}}}
         %\caption{Optimal APD.}
         %\label{fig:five over x}
     \end{subfigure}
        \caption{Generating an artificial microstructure (see Section \ref{sec:ebsd-synthetic}). (A): A scatter matrix plot of the artificially generated data, sampled from a distribution fitted to the EBSD data in Figure~\ref{fig:ebsd_fit1}, with colours representing density (high in bright colours, low in dark). (B): The resulting optimal APD (with tolerance $\varepsilon = 0.02 = 2\%$) was obtained with  Algorithm~\ref{alg:Lloyds} with 5 iterations of Lloyd's algorithm and the initial seed points $X^0$ drawn randomly from the uniform distribution on $\Omega$. This 
        %(see Algorithm~\ref{alg:Lloyds}), which 
        took \SI{266}{\second} to generate. In the absence of crystallographic input, the colours are assigned randomly. To compare the artificial sample with the real EBSD data, we have run the two-sample Kolmogorov-Smirnov test for marginal distributions. The returned $p$-values of $0.973$ (grain areas), $0.968$ (ellipse aspect ratios) and $0.706$ (ellipse angles) show a high degree of statistical equivalence.}
        \label{fig:ebsd_artificial1}
\end{figure}

Following the fitting, we sample $(v_i,a_i/b_i,\theta_i)$ from the resulting joint distribution and use these to construct the anisotropy matrices $\mathbf{A}_i=\mathbf{A}(\hat{a}_i,\hat{b}_i,\theta_i)$; see equation \eqref{eq: A}. Then we
employ Algorithm \ref{alg:Lloyds} to obtain an optimal APD with grains of volume $v_i$. The reason for using Algorithm~\ref{alg:Lloyds} is to generate more `regular' APDs; the algorithm has the effect of significantly reducing the number of disconnected grains and non-simply connected grains, or eliminating them altogether. (Note that Algorithm \ref{alg:Lloyds} does not require any input for the seeds $x_i$.) 
%{\db \sout{with prescribed properties and in which grains are not disjoint.}}
%\cdb{`with prescribed properties' is a bit vague. Also, why do we mention that the grains are not disjoint? They never are; they always intersect along their boundaries.}
%\cmb{What I meant was that no single grain is disjoint, meaning it does not have two or more disconnected components. If we do not run Lloyds, then we see such disjointness / disconnectedness}
To assess whether the artificial sample is statistically equivalent to the real EBSD data, we perform the two-sample Kolmogorov-Smirnov test for marginal distributions using \textsc{OpenTurns} \cite{OpenTurns}. The results are presented in Figure~\ref{fig:ebsd_artificial1}.
%{\ks COMMENT: This is great! In many problems, like ML problems or investigating the effect of uncertainties in the microstructure, we need to generate a large number of RVEs. Your approach speeds up and simplifies the procedure to generate microstructure with specified variations and uncertainties. }

\subsection{Modelling challenging geometries}\label{sec:ex_3d_printed_steel}
%In the last example, we aim to showcase the versatility of the method with respect to grain geometries that can be produced.

%Firstly, in a nod to our industrial partner, Tata Steel, whose EBSD measurements we rely on, in this example we create a 2D APD with cells aligning to recreate its logo. The results are presented in Figure~\ref{fig:Tata_logo}. 

%Inspired by \cite[Section~3.3]{KCLW22}, in order to showcase the versatility of the method with respect to {\db the} grain geometries that can be produced, in the final example we {\db \sout{aim to}} create a multi-phase optimal APD imitating 
%{\db the} polycrystalline grain texture of an additively manufactured material and compare it with EBSD measurement of a specimen from bidirectionally-printed single-track thickness 316L stainless steel wall built by directed energy deposition, taken directly from \cite{BCC20}. 
%The results are presented in Figure~\ref{fig:AM_example}.

\begin{figure}[htb]
    \centering
    \begin{subfigure}[t]{0.68\textwidth}
        \centering
        \subcaption{}
        \includegraphics[trim=0 17 0 0,clip,width=\textwidth]{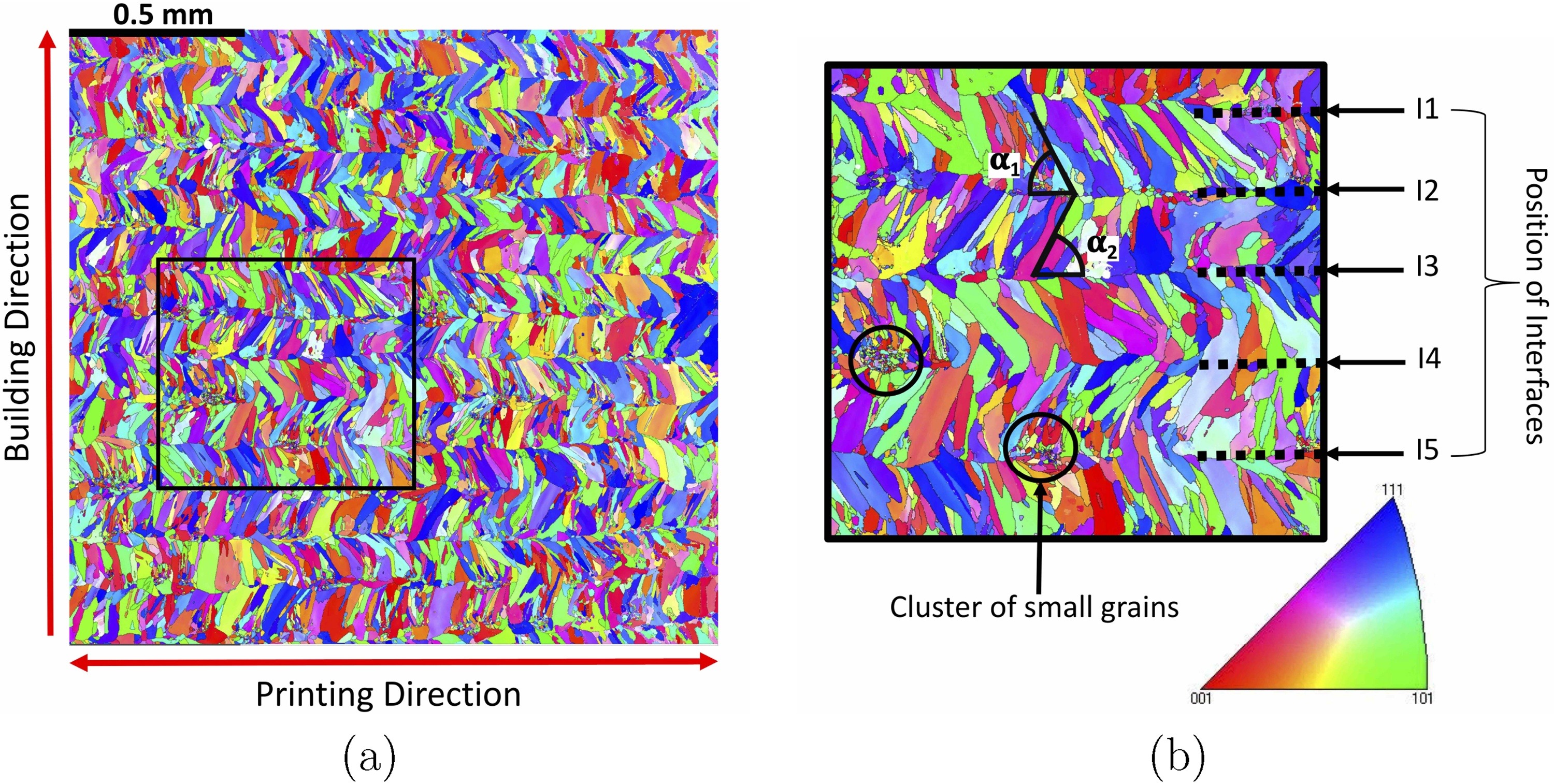}
        %\caption{Tata Steel logo with ellipses.}
        %\label{fig:three sin x}
    \end{subfigure}
    \hfill
    \begin{subfigure}[t]{0.28\textwidth}
        \centering
        \subcaption{}
        \includegraphics[trim=110 100 160 150,clip,width=\textwidth]{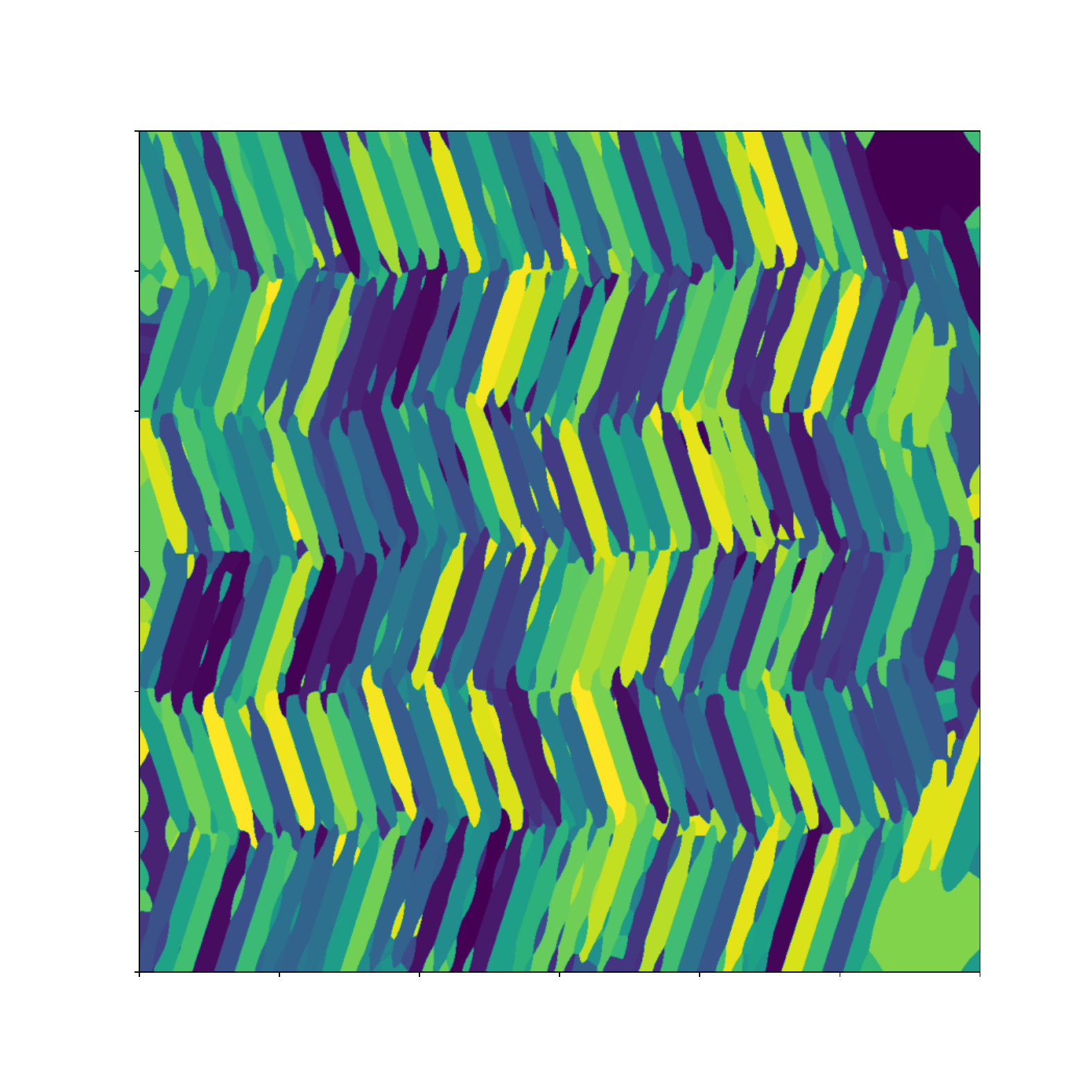}
        %\caption{Tata Steel logo as an APD.}
        %\label{fig:five over x}
    \end{subfigure}
       \caption{(A):~The IPF color map parallel to the building direction for the EBSD scan of a bidirectionally-printed single-track thickness 316L stainless steel wall,
       %{\db \sout{taken from \cite{BCC20},}}
       with the inset zooming in on 5 phases. This figure is taken directly from \cite{BCC20}. (B):~An example APD imitating such a 5-phase geometry. It is generated by supplying anisotropy matrices corresponding to thin ellipses tilted at prescribed angles, and centroids and volumes ensuring that the ellipses fill the space well. We also add several random small grains and run a few iterations Lloyds algorithm (see Algorithm~\ref{alg:Lloyds}). The overall runtime to generate it was \SI{3}{\second}.}
       \label{fig:AM_example}
\end{figure}

%\cdb{I have cut up the previous version of the following into several sentences, since originally the sentence was a bit long (7 lines).}
 In our final example, inspired by \cite[Section~3.3]{KCLW22}, to showcase the versatility of the method with respect to the grain geometries that it can produce, we create a highly-anisotropic APD imitating 
a 3D-printed stainless steel. This 
additively-manufactured material 
from \cite{BCC20} is a  bidirectionally-printed single-track thickness 316L stainless steel wall, built by directed energy deposition. The results are presented in Figure~\ref{fig:AM_example}.

\section{Discussion}
In this section we will discuss how our methods compare and fit in with the existing body of literature and provide an outlook about future work. 

Several interesting APD-based approaches to the modelling of microstructure in metals have been introduced and explored in recent years. Starting with \cite{ABGLP15}, and 
%in more recent 
more recently in
\cite{AFGK23}, the authors propose various techniques for converting an EBSD data grain map, which assigns each pixel to a grain, into an APD in such a way that the number of misassigned pixels is minimised. 
The control over the area/volume of the APD grains is introduced via approximate \emph{weight-constraints}, and the 
%overall
resulting optimisation problem
%algorithm 
is a linear programming 
%one
problem. Still in the realm of trying to fit an APD directly to pixel-level data, authors in \cite{VBWBSJ16} propose a fast stochastic optimisation-based alternative to the linear programming approaches {
%\db \sout{championed}}
introduced
in \cite{ABGLP15}. Yet another approach to direct fitting is the so-called \emph{gradient descent-based tessellation fitting} introduced in \cite{FPFDUSS21} in the broader context of generating realistic artificial Li-ion electrode particle architectures, and extended to APDs in \cite{PFWKS21}. 
%{\db Finally, \cite{TR18} propose a heuristic for fitting APDs to grain maps that avoids solving an optimisation problem altogether, which is very fast but yields a higher misassignment error.} 

%\cdb{Do we cite the Neper and Dream3D software anywhere? Maybe it comes later, or maybe I missed it. Anyway, I feel like we should cite them, although perhaps this is the wrong section since they don't use APDs. Perhaps in the introduction?} \cmb{Very good point and we currently do not, I will add it!}

All of the methods mentioned so far focus on minimising the number of misassigned pixels.  In the example presented in Section~\ref{sec:ex_ebsd} we present an alternative approach of first post-processing the EBSD pixel data to obtain a grain file, followed by employing Algorithm~\ref{alg:optimal_APD} to minimise the area/volume error. It does not explicitly focus on the pixel-level mismatch, but  nonetheless seems to yield
a similar (but slightly lower) level of accuracy. On the other hand,}
%similarly accurate results  
%At the same time, 
% \cdb{Too many "At the same times" - it also comes in the following sentence.}
we benefit from the precise fitting of the volumes, and also, thanks to the GPU-friendly implementation, from a much decreased runtime. 
%{\mb At the same time, since variants of Algorithm~\ref{alg:pixel_APD} lie at the heart of {\mb \sout{all} many} the APD-based approaches, we believe that other methods present in literature can benefit from our GPU-friendly implementation of this algorithm and can expect a near three orders of magnitude speed-up against a baseline CPU implementation, as we reported in Figure~\ref{fig:runtime-apd-gen}. }
%\cdb{This is perhaps overselling it a bit? A method will only see the three-orders of magnitude speedup if the bottleneck is computing the diagrams. It could be that the runtime is dominated by something else, e.g., by the linear programming solver or by the stochastic optimisation algorithm. I would tone this down a bit, perhaps something like the following:}
%\cmb{This is a fair point, although if really the super slow computation of APDs is not a botttleneck of some approach, then that approach must be so so slow. I guess the old wording would be fine when caveated with "for large enough $M$ and $N$, as I guess eventually the computation of APD would take over as a bottleneck, but the new wording is perfectly fine with me.}
At the same time, any other approach in the literature where APDs need to be computed may benefit from our GPU-friendly-implementation of Algorithm~\ref{alg:pixel_APD}, which computes APDs three orders of magnitude faster than a baseline CPU implementation, as we report in Figure~\ref{fig:runtime-apd-gen}.

Another set of methods
%{\db \sout{explored}}
in literature 
for fitting APDs to EBSD data uses 
a heuristic guess for $(X,W,\La)$ and 
avoids solving an optimisation problem altogether; see \cite{TR18,AFGK23B}.
%concerns refraining from finding the optimal APD altogether 
%\cdb{It's not quite accurate to say optimal APD because also the methods of Alpers et al. don't find optimal APDs in the sense that we mean in this paper}
%and instead approximating 
%it with an explicitly defined heuristic guess for the set of optimal weights $W$, given $X$ and $\La$, see \cite{TR18,AFGK23B}. 
This carries next to no computational cost  and gives a similar (but slightly lower) pixel misassigment error. However, 
%but, 
as we have reported in Section~\ref{sec:ex_ebsd}, the heuristic guess for $W$ from \cite{TR18}
%{\db \sout{, which is included in our library,}}
appears to struggle to get the volumes of the grains right. 
%\cdb{I wouldn't mention our library here; it makes it sound like our library may struggle to get the volumes right.} 
On the other hand, using this heuristic as an initial guess seem to reliably improve the runtime of Algorithm~\ref{alg:optimal_APD} when employed to fit an optimal APD to real EBSD data.

We also wish to mention the work \cite{vNvDG21}, in which the authors considered a grain growth model using 
%an 
anisotropic Voronoi diagrams (weights %W$ 
%always
$w_i$ all
equal to zero).
%growth model. 
In light of the recent work on grain growth models using APDs in \cite{AFGK23B},
%using the heuristic guess $W= 0$,} 
%heuristic guess with $W= 0$ in \cite{AFGK23B},
%based on adjusting $X$ and $\La$, 
%\cdb{I don't really know that this means - I think it would be better to say a bit more or a bit less, e.g., delete the phrase ``using the heuristic guess $W= 0$,
%based on adjusting $X$ and $\La$"}
%\cmb{I agree and probably better to say less at this point}
it would be interesting to see how our optimal APDs can be 
%{\db \sout{compared}}
used to infer quantities such as the growth velocity of each nucleated grain.

%\cmb{New wording below, I guess David you will say that this is too confrontational, so please adjust :) But I just wanted to bring this even for our internal discourse, as I dug a little bit more into how this is done in Neper.}

An area in which we think our method is %{\db \sout{uniquely}} 
useful 
%concerns 
is the
reliable generation of samples of realistic synthetic microstructures with prescribed statistical properties,
%concerning 
such as
grain sizes and anisotropy. Existing approaches, such as \textsc{Dream3D} \cite{groeber2014dream}, \textsc{Neper} \cite{Quey2011} and \textsc{Kanapy} \cite{Biswas2020} use (isotropic) power-diagram-based algorithms. Since in any power diagram the spatial anisotropy of grains is determined primarily by the relative location of seed points of neighbouring grains %\cdb{the weights also play a role}, 
%substantial iterative optimisation algorithms are
many iterations of an optimisation algorithm may be
needed to produce a power diagram with a desired distribution of anisotropy. %{\db \sout{among the generated sample.}}
In the case of \textsc{Neper}, it is reported in \cite{QR18} that it took about 4.8 hours to generate a sample with $N=10,000$ grains in 3D with prescribed anisotropy of grains.  Based on the timings reported in Figure~\ref{fig:runtime-optimal-apds}, our method is expected to do so in about 5-10 minutes. (Note, however, that the runtimes in \cite{QR18} were produced using CPU hardware from six years ago, so this is not an entirely fair comparison.) 
%\cdb{Just to try to soften the blow a bit!}
An alternative approach is to employ deep-learning tools (GANs) to generate artificial microstructures, as done in \cite{chun2020deep} and in \textsc{DRAGen} \cite{henrich2023dragen}. We further note that there is interest 
%for
in
this task in biology \cite{Song22}.  %As reported Typical tessellation optimization involves a large number of iterations, of the order of 105–106 for N = 1,000. At
%each iteration, the tessellation and its associated objective function must be computed

%As already discussed, power diagrams   and, to have better control over anisotropy of grains, require clever iterative  combined with 

%There is interest for this task in biology \cite{Song22},
%but to the best of our knowledge, no other method present in the literature is able to achieve this in reasonable time for metallic microstructures. {\mb [soften it a bit]} 

Moving forward, we would like to accelerate our methods further 
%through 
by
employing adaptable pixel/voxel sizes. We believe that tools for identifying pixels/voxels at the boundary of a grain developed in \cite{SWKKCS18} will prove useful in this regard. There are some similar ideas in the optimal transport community too \cite{dieci2019boundary}. We expect the idea of coresets developed in \cite{AFGK23} to be similarly helpful in addressing this challenge. Our library already allows users to manually supply a non-uniform discretisation of the domain $\Omega$, but it is key for such a procedure to be automated. A local refinement of the discretisation can be implemented in a GPU-friendly way by employing the idea of masking.

%{\mb Yet another alternative to finding an optimal APD by minimising either pixel mismatch or grain volume deviations, is to . }

%This is taken further in \cite{SWKKCS18}, where notably authors introduced methods for identifying . We see this as potentially crucial to further accelerating our methods through employing adaptable pixel/voxel sizes. 

%As already discussed, in  and, more recently, in \cite{AFGK23B}, the authors discuss heuristic guesses for the set of optimal weights $W$, given $X$ and $\La$. As already mentioned, the heuristic guess from \cite{TR18} is included in our library.

As demonstrated in Section~\ref{sec:ebsd-synthetic}, our method exhibits remarkable time efficiency in generating realistic RVEs with a large number of grains and authentic morphology. This capability opens the door to systematically generating a large number of carefully designed RVEs, particularly required for machine learning and data science applications. The demand for creating a substantial quantity of representative microstructures is fundamental for conducting comprehensive studies in these fields.

%{\db \sout{Finally,}}  
Given the speed 
%{\db \sout{and efficiency}} \cdb{Is there a diffrence between speed and efficiency?}
of our method and recent work on employing machine-learning tools to learn the evolution of a two-phase microstructure \cite{OSGDK22}, 
%{\db \sout{it is our aim to explore the idea of studying}}
we plan to study
the evolution of a microstructure in steel under deformation, as done recently in \cite{STRSRD22}.
%, where 
%\cdb{The sentence was 6 lines long, so I cut it in two.}
The evolution of the microstructure %{\db \sout{is cast}}
could be modelled
as a time evolution of %the 
an
optimal APD, generated by 
%$\{X^t, \La^t, \overline {W}^t\}$ 
$(X^t,W^t,\La^t)$ 
%\cdb{I changed the notation to be consistent with what we use earlier. I also removed the overline on $W^t$. I see why you included it, to denote the optimal weights rather than any set of weights, but we've not used this notation anywhere else and I don't think we need to be so precise here.}
and the set of target volumes $(V^t)$, where $t$ denotes time. 

Finally, apart from the applications in microstructure modelling, we mention that it would be easy to modify our library \textsc{PyAPD} \cite{PyAPD}
to solve quite general semi-discrete optimal transport problems with non-quadratic costs, which might be beneficial to the optimal transport community. Currently our code is limited to the anisotropic transport cost $c(y,x_i)=| y - x_i |_{\mathbf{A}_i}^2$, but it could be modified to work for any cost that can be represented in \textsc{PyKeOps} \cite{CFGCD21} as a LazyTensor.

\section*{Acknowledgements}
DB and MB would like to acknowledge the support of the Engineering and Physical Sciences Research Council in the UK, as part of the grant EP/V00204X/1  Mathematical Theory of Polycrystalline Materials.
Part of this research was performed while MB was a visiting fellow at the Institute for Pure and Applied Mathematics (IPAM), as part of the long program \emph{New Mathematics for the Exascale: Applications to Materials Science}. IPAM is supported by the U.S. National Science Foundation (Grant No.~DMS-1925919). As a result, this research used resources of the National Energy Research Scientific Computing Center (NERSC), a U.S. Department of Energy Office of Science User Facility located at Lawrence Berkeley National Laboratory, operated under Contract No.~DE-AC02-05CH11231 using NERSC award DDR-ERCAP0025579.

%\newpage

%\medskip

\printbibliography

\end{document}